\documentclass[
aps,%
12pt,%
final,%
notitlepage,%
oneside,%
onecolumn,%
nobibnotes,%
nofootinbib,
superscriptaddress,%
noshowpacs,%
centertags]%
{revtex4}

\begin{document}
\selectlanguage{english} 

\begin{flushleft}
Accepted for publication in Astronomy Reports
\end{flushleft}

\title{Determining the Parameters of Massive Protostellar Clouds
via Radiative Transfer Modeling}

\author{\firstname{Ya. N.} \surname{Pavlyuchenkov}}
\affiliation{Institute of Astronomy, Russian Academy of Sciences, Moscow, Russia}

\author{\firstname{D. S.} \surname{Wiebe}}
\affiliation{Institute of Astronomy, Russian Academy of Sciences, Moscow, Russia}

\author{\firstname{A. M.} \surname{Fateeva}}
\affiliation{Institute of Astronomy, Russian Academy of Sciences, Moscow, Russia}

\author{\firstname{T. S.} \surname{Vasyunina}}
\affiliation{Max-Planck Institute for Astronomy, Heidelberg, Germany}

\begin{abstract}
A one-dimensional method for reconstructing the structure of prestellar and protostellar clouds
is presented. The method is based on radiative transfer computations and a comparison of theoretical
and observed intensity distributions at both millimeter and infrared wavelengths.
The radiative transfer of dust emission is modeled for specified parameters of the density
distribution, central star, and external background, and the theoretical distribution of
the dust temperature inside the cloud is determined. The intensity distributions at millimeter
and IR wavelengths are computed and quantitatively compared with observational data.
The best-fit model parameters are determined using a genetic minimization algorithm, which
makes it possible to reveal the ranges of parameter degeneracy as well. The method is illustrated
by modeling the structure of the two infrared dark clouds IRDC-320.27+029 (P2)
and IRDC-321.73+005 (P2). The derived density and temperature distributions can be used to model
the chemical structure and spectral maps in molecular lines.
\end{abstract}

\maketitle

\section{INTRODUCTION}

Massive stars are one of the most important
components of our Galaxy and other galaxies, since
they determine the energy balance in the interstellar
medium and are the principal source of the heavy
elements. However, the formation of massive stars
is understood less well than the formation of low-mass
stars ($M<10\,M_\odot$). It is not yet known whether
massive stars form like low-mass solar-type stars
or stars with different masses form in different ways
(see \cite{Zinnecker}).

The general scenario of solar-type star formation
has been developed with a fairly high degree of confidence.
It starts with the formation of a dense collapsing
core in a molecular cloud. During the core collapse,
a central protostellar object and circumstellar
torus appear, which are later transformed into a young
star and a gas–dust disk, respectively. The durations
of the various stages of this process and the details of
the transitions between them remain open problems.
However, objects at different stages of star formation
have long been observed: prestellar (starless) cores,
class 0 and I protostars, classical T Tauri stars (class
II protostars) and weak-lined T Tauri stars (class III
protostars).

No similar sequence of formation stages has been
developed for massive stars, as a result of both observational
and theoretical difficulties. All regions of
massive star formation are far from us, and so require
observations with high angular resolution (the distance
of the closest, in Orion, is approximately 500 pc;
typical distances are of the order of 2 kpc or more).
The high gas densities in these regions lead to high
extinctions, making optical and even IR observations
difficult. As the energy of massive stars influences
the surrounding gas starting from the earliest stages
of their existence, regions of massive star formation
have very complex structure, which requires the application
of three-dimensional models in theoretical studies.

The detection of massive prestellar cores --- the
counterparts of low-mass prestellar cores --- could be
an important step toward understanding the nature
of massive star formation. We expect them to have
the simplest structure, while still providing important
information concerning the initial conditions for massive
star formation. The most promising candidate
objects are Infrared Dark Clouds (IRDCs). These
clouds are seen in absorption against the Galactic
IR background over wavelengths from several $\mu$m to
several tens of $\mu$m; they were detected as a result of
surveys with the ISO \cite{ref1} and MSX \cite{ref2}
space telescopes. Their cores, i.e., the densest IRDC regions,
can also be detected in emission at millimeter and
submillimeter wavelengths. Some of these demonstrate
signs of star formation \cite{ref3}, but the crucial test
will be the detection of massive {\em starless} cores, i.e.,
without embedded compact sources and with collapse
signature in their spectra.

This makes searches for molecules whose spectral
transitions could trace the kinematics of massive
prestellar cores topical. Finding such molecules and
transitions requires modeling the chemical evolution
and radiative transfer based on available information
about the distributions of the gas and dust densities
and temperatures in these objects. In the case of
low-mass cores, information about their structure is
obtained from data on either their emission in the
submillimeter and millimeter \cite{launhardt}, or the absorption of
the background stellar emission \cite{alves} in the optical and
NIR.

Spectra of infrared dark clouds can be studied in
more detail. First, due to the presence of the infrared
background, they can be observed in absorption over
a much broader wavelength range than typical clouds
in regions of low-mass star formation. Second, like
other gas–dust clouds, IRDC cores can be observed
both in absorption and emission, making it possible
to construct their spectral energy distributions
(SEDs) in more detail, and, consequently, to reproduce
their physical structure more reliably. However,
this requires detailed numerical modeling and is a fairly
resource-intensive task.

In this paper, we present a method for studying
the density and temperature distributions in prestellar
cores, based on modeling the radiative transfer in
them taking into account both their own radiation and
the absorption of the background radiation at NIR to
millimeter wavelengths. As an example, we use two
dense IRDC cores from the sample of Vasyunina et
al. \cite{vasyuninaetal2009}, IRDC 320.23+0.32 and IRDC 321.71+0.07.
The observational data at 1.2 mm described in 
\cite{vasyuninaetal2009}\footnote {Available at the Strasbourg
astronomical Data Center (CDS) at http://cdsweb.u-strasbg.fr/cgi-bin/qcat?J/A+A/499/149}
are available for these cores, as well as Spitzer maps
at wavelengths from 3.5 to 70 $\mu$m, obtained as a result
of the GLIMPSE \cite{glimpse} and MIPSGAL \cite{mipsgal} surveys.

\begin{figure}[!t]
\setcaptionmargin{5mm}
\onelinecaptionsfalse
\captionstyle{normal}
\includegraphics[scale=0.35,clip=]{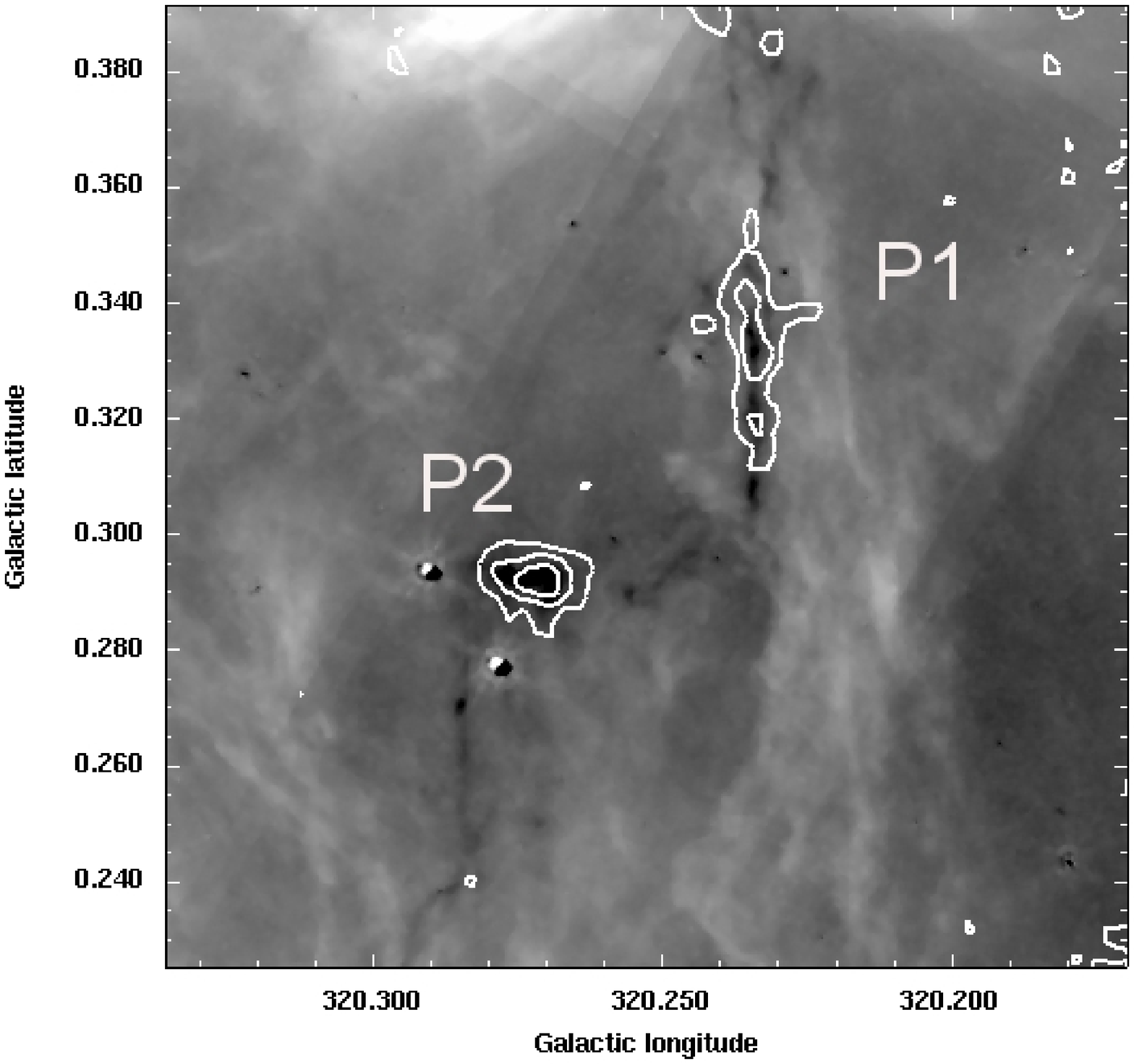}
\includegraphics[scale=0.35,clip=]{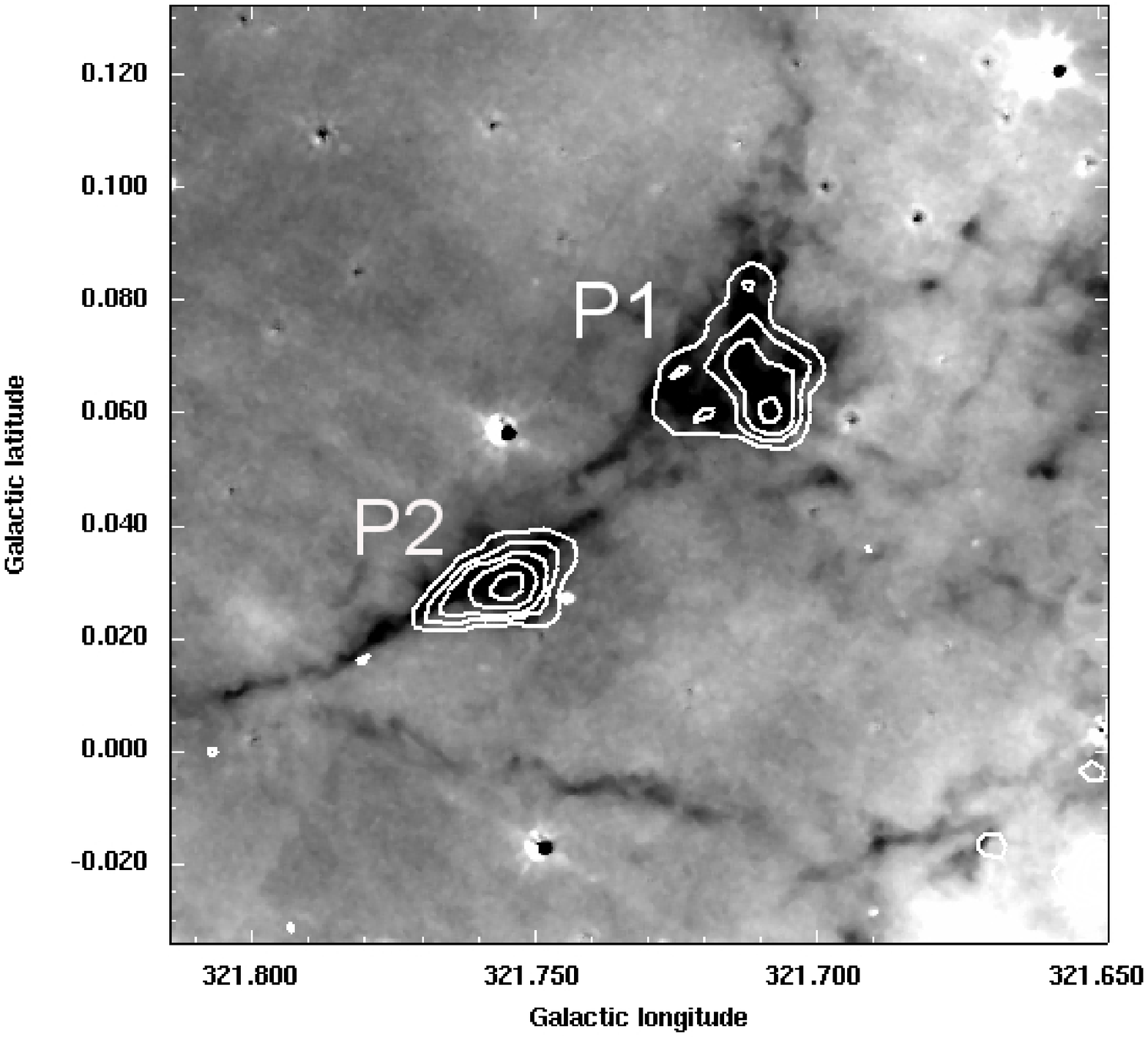}

\includegraphics[scale=0.3,clip=]{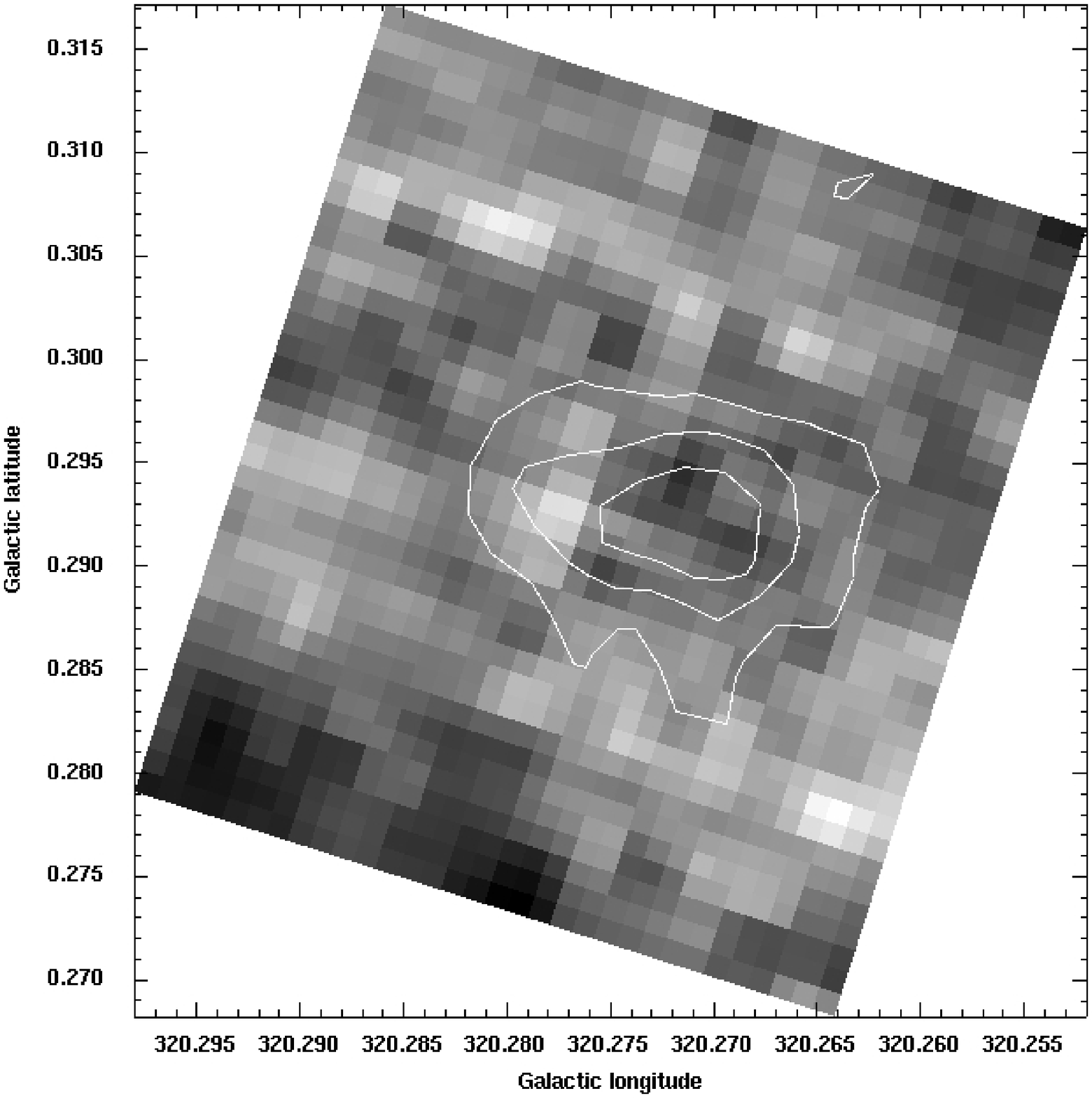}
\includegraphics[scale=0.3,clip=]{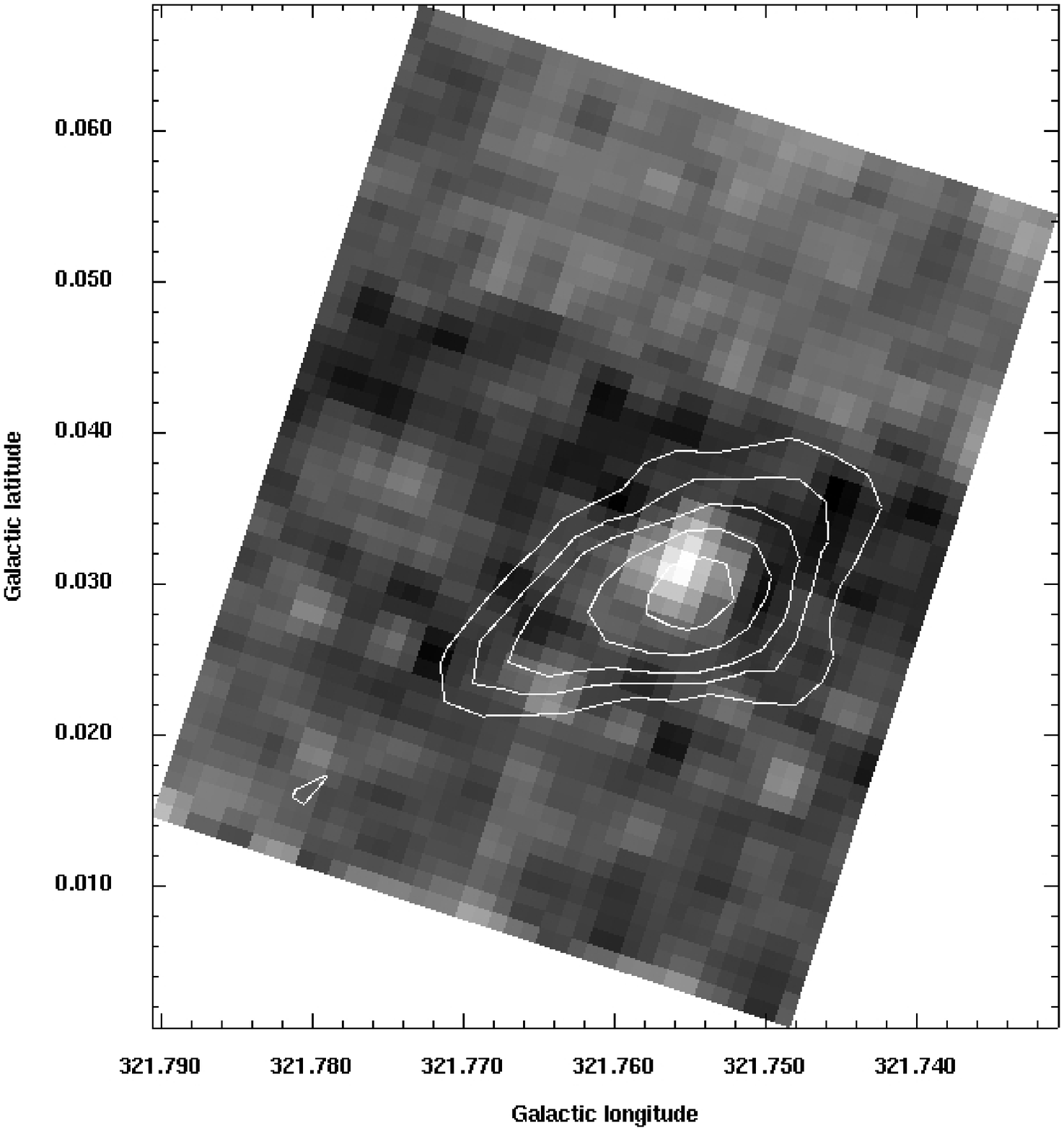}
\caption{
Intensity distributions in the IRDC 320 (left column) and
IRDC 321 (right column) at 8 $\mu$m (top row) and 70 $\mu$m
(bottom row). The grey scale shows the IR emission and the contours
show the emission at 1.2 mm.}
\label{maps}
\end{figure}

We selected the sources P2 in IRDC 320 and
P2 in IRDC 321 for our study. Both these sources
appear round in millimeter maps (Fig. \ref{maps}), suggesting
that a spherically symmetric approximation
may be applicable. There is an important difference
between the sources: 70 $\mu$m emission is associated
with IRDS 321, whereas IRDS 320 is not detected
in either emission or absorption at 70 $\mu$m. If we are
dealing with massive prestellar or protostellar cores,
the presence of emission at 70 $\mu$m may indicate a
later evolutionary stage. For brevity, we further
call these cores IRDC 320 and IRDC 321. Table~\ref{obsdata}
presents their coordinates, distances, and the masses
and column densities derived from millimeter observations.

The special feature of the method used here is that
we reproduce the observational data at millimeter,
near-IR, and mid-IR wavelengths in the framework of
a single model. Similar modeling had been performed
earlier, but using smaller numbers of wavebands. In
particular, observations at 450 and 850 $\mu$m were used
to determine the thermal structure of an IRDC core
seen toward the giant molecular cloud W51 \cite{ormel}.
However, as we will show below, the use of only long wavelength
data may be insufficient for this problem.

\begin{table}
\caption{}
\label{obsdata}
{\bf Parameters of the modeled IRDC cores from \protect{\cite{vasyuninaetal2009}}.}
\begin{tabular}{lllccc}
\hline
Source & $\alpha$ (J2000.0) & $\delta$ (J2000.0) & Distance, & Mass (1.2 mm), & Column density, \\
  & & & kpc & $M_\odot$& $10^{22}$ cm$^{-2}$ \\
\hline
IRDC 320.23+0.32 (P2) & 15$^{\rm h}$ 07$^{\rm m}$ 56.7$^{\rm s}$ & --57$^\circ$ 54$^{\prime}$ 27$^{\prime\prime}$ & 1.97 & 50 & 1.5 \\
IRDC 321.71+0.07 (P2) & 15$^{\rm h}$ 18$^{\rm m}$ 26.7$^{\rm s}$ & --57$^\circ$ 21$^{\prime}$ 56$^{\prime\prime}$ & 2.14 & 110 & 3.2 \\
\hline
\end{tabular}
\end{table}

\section{SEARCH FOR BEST-FIT CLOUD PARAMETERS}

The reconstruction of a source structure based on
observational data is an inverse problem. Such problems
are usually ill-defined and cannot be solved
without some initial assumptions about the structure
of the studied region. Usually, a specified model with
several free parameters is used, which can be found by
fitting the model to the observational data. When the
number of free parameters is large, some optimization
algorithm must be used to find the best fit. Such algorithms
often implement a search for the minimum of a
function of several variables, where some criterion for
the agreement between the model and observations
serves as this function and the relevant free parameters
are the variables. Here, we searched for the best-fit
parameters applying the PIKAIA numerical code~\cite{pikaia},
which uses a genetic algorithm. This code yields
both the localization of the functional minimum in the
space of several parameters and the degeneracy of the
parameters. Note that PIKAIA has already been used
in some astrophysical studies (see.,
e.g.\cite{pikaia_agb,pikaia_proplyd}).

\subsection{Example of Applying the Method:
A Two--Component Model}

We illustrate the method used to search for the
best fit using the example of a simple two-component
model of a protostellar cloud; various modifications
of this procedure are fairly popular when analyzing
observational data~\cite{2layer,robitaille}.
We suppose that the cloud consists of two components,
each with its own dust temperature $T$ and dust surface
density $\Sigma$. Physically these two components may be,
for example, the cloud core and an envelope, but their
physical interpretation does not matter from the computational
point of view. For definiteness, we consider the first component to
be located behind the second. The radiation intensities
of the first and second components propagating toward the
observer are then given by
\begin{align}
&I_\text{1}(\nu) = \left(1-e^{-\kappa_{\nu}\Sigma_\text{1}}\right)B_{\nu}(T_\text{1})\\
&I_\text{2}(\nu) = e^{-\kappa_{\nu}\Sigma_{2}}\,I_\text{1}(\nu)
+\left(1-e^{-\kappa_{\nu}\Sigma_\text{2}}\right)B_{\nu}(T_\text{2}),
\end{align}
where $\kappa_{\nu}$ is the dust absorption coefficient per unit
mass of dust and $I_\text{2}(\nu)$ is the intensity received by the
observer. These formulas neglect scattering of the
light, and take into account only the intrinsic thermal
emission of the medium. Thus, the radiation intensity
in the two-component model depends on the four
parameters $T_\text{1}$, $T_\text{2}$, $\Sigma_\text{1}$
and $\Sigma_\text{2}$. For convenience, the
surface densities $\Sigma_\text{1}$ and $\Sigma_\text{2}$
can be converted into molecular-hydrogen column densities 
$N_1$ and $N_2$ assuming
the dust-to-gas mass ratio to be 0.01~\cite{bochkarev}.
The model will be compared with the SED observed
over six wavelengths (1.2 mm, 70, 24, 8, 5.8, and
3.6 $\mu$m), using the observed intensities toward the
assumed center of the cloud, determined as the peak
of the 1.2-mm emission. The chosen criterion for
agreement between the model and observations takes
the form
\begin{equation}
\chi^2 = \dfrac{1}{N}\sum\limits_{i=1}^{N} \left(\lg I^{\rm obs}_i - \lg I^{\rm mod}_i\right)^2/\sigma_i^2,
\end{equation}
where $I^{\rm obs}_i$ and $I^{\rm mod}_i$ are the observed
and theoretical radiation intensities for the $i$th frequency channel and
$\sigma_i$ is the dispersion of the logarithm of the observed intensity.

\begin{figure}[!t]
\setcaptionmargin{5mm}
\onelinecaptionsfalse
\captionstyle{normal}
\includegraphics[scale=0.62]{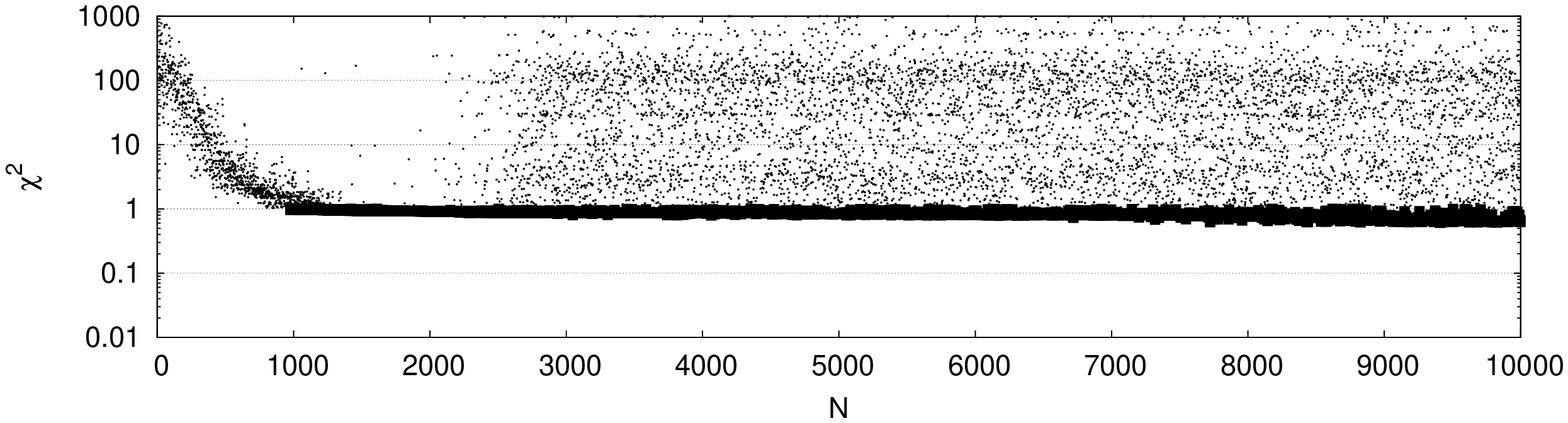}
\includegraphics[scale=0.55]{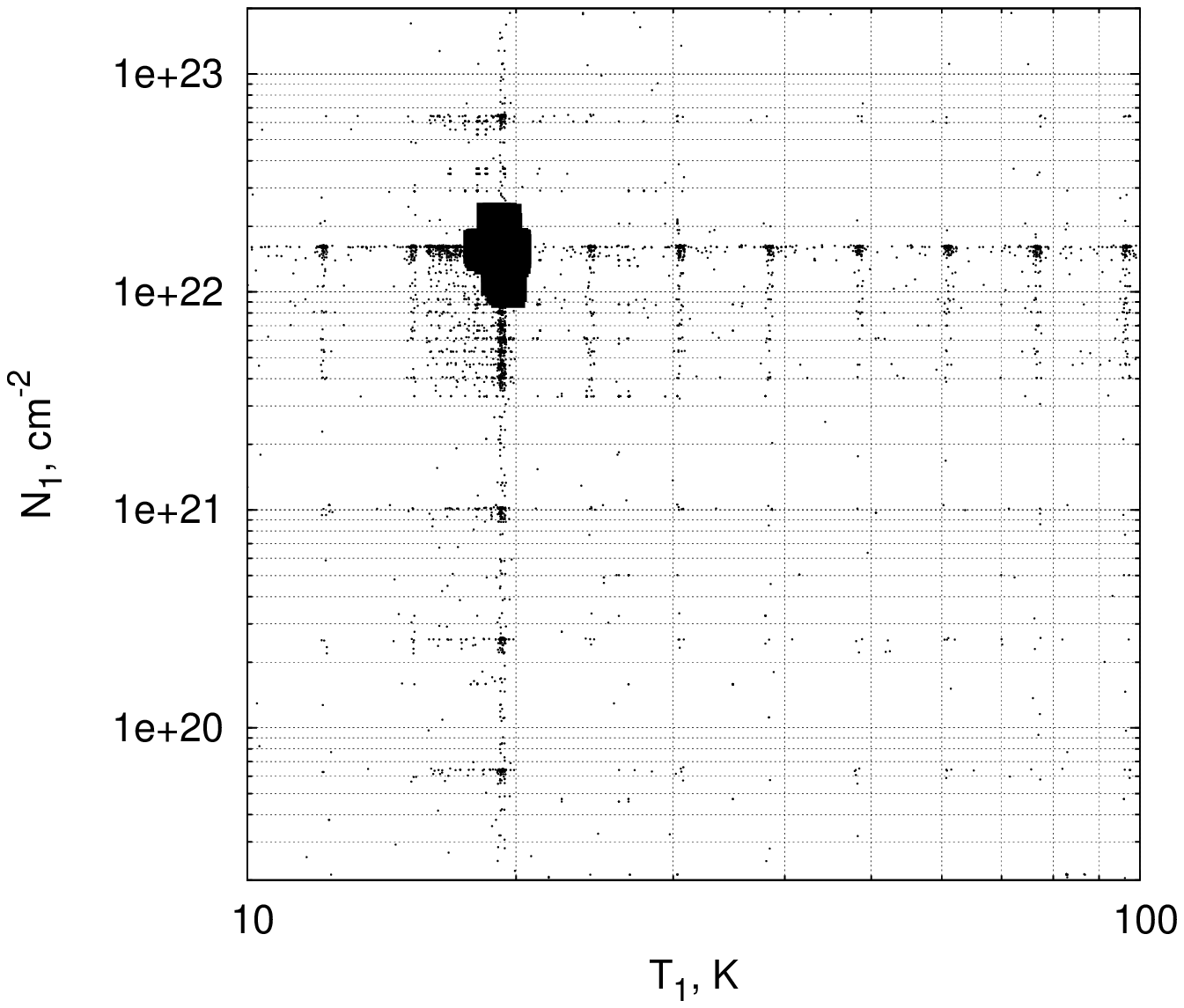}
\includegraphics[scale=0.55]{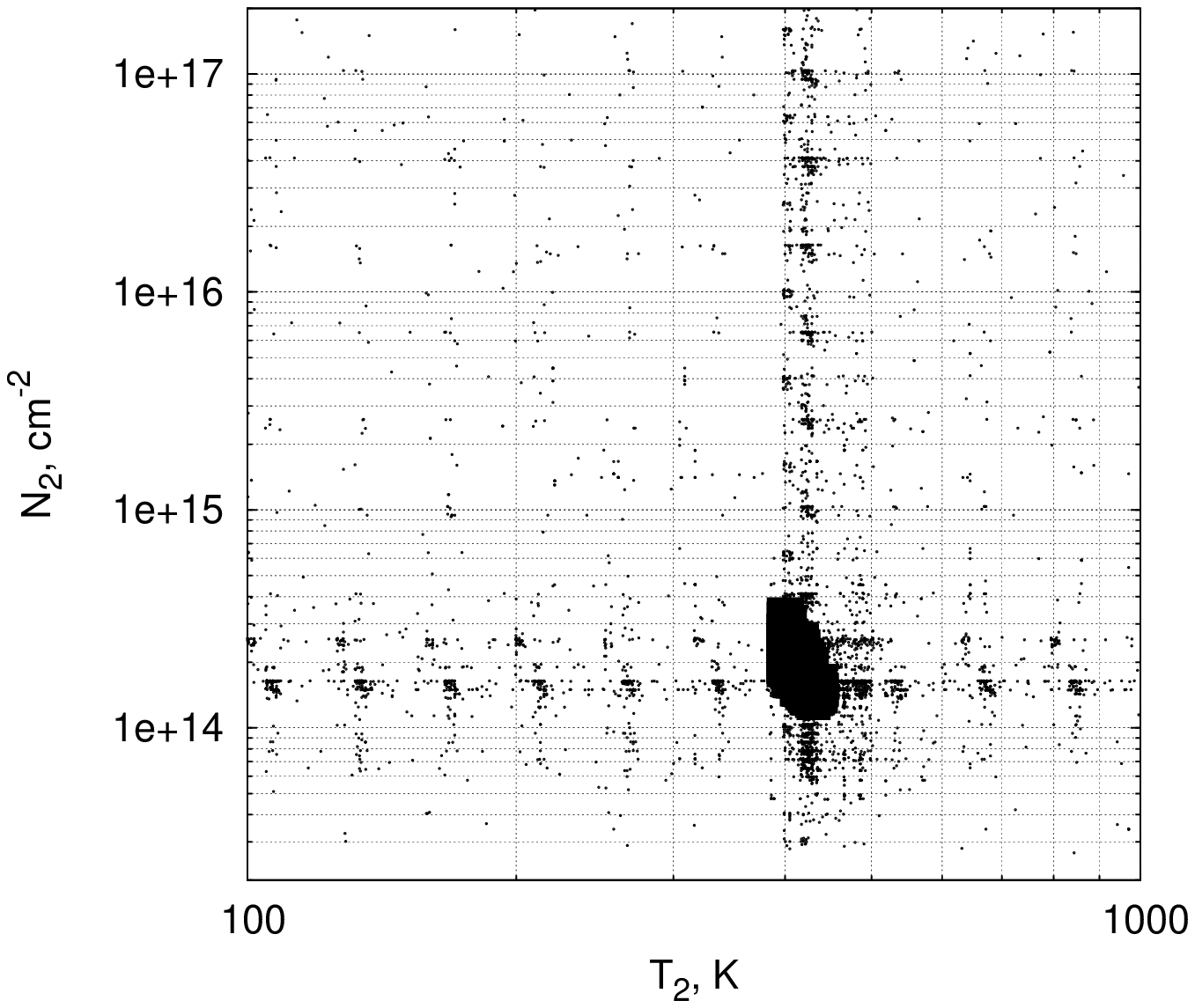}
\caption{Results of searching for best-fit parameters using the two-component
model for the protostellar cloud IRDC 321. The upper plot shows the history of the model’s
convergence. The lower left plot demonstrates the localizations of the model
parameters in $T_1$--$N_1$ space, and the lower right plot shows the same in $T_2$--$N_2$ space.
The black squares show models with $\chi^2<1$ and the dots show other models.}
\label{pic0a}
\end{figure}

\begin{figure}[!t]
\setcaptionmargin{5mm}
\onelinecaptionsfalse
\captionstyle{normal}
\includegraphics[scale=0.6]{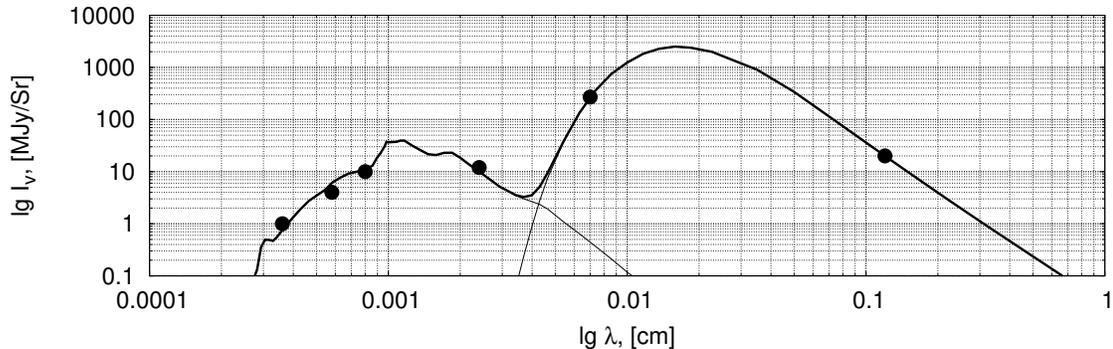}
\caption{Comparison of the SED of the best-fit two-component model with observations
of IRDC~321. The thin curves demonstrate the contributions of the first and second
components, and the thick curve the total SED. The circles show the
observed intensities.}
\label{pic0b}
\end{figure}

Figure~\ref{pic0a} shows the results of the minimization
based on the PIKAIA algorithm. The computation
begins with arbitrary parameter values (only their
limits are specified); therefore, there is initially
poor agreement with the observations, i.e., the $\chi^2$
values are very large (Fig.~\ref{pic0a}, upper plot). After calculating
about a thousand models, the algorithm finds a
minimum with $\chi^2<1$; however, the search for the solution
continues, since the algorithm checks whether
the minimum is local or not. Further, there are no
changes, even after computing 10 000 models. The
lower plots in Fig.~\ref{pic0a} show the locations of the model
parameters in the $T_1$--$N_1$ and $T_2$--$N_2$ spaces. The
black squares represent the models with $\chi^2<1$, i.e.,
models that correspond to the minimum in the upper
plot of Fig.~\ref{pic0a}. Figure~\ref{pic0b} shows a comparison of
the best-fit model with the observational data for
IRDC~321. The model parameters are sharply localized
and provide a good agreement between the
theoretical SED and the observed points.

Formally, the comparison of the model with the
observations shows that the real cloud can be represented
by a cold core with temperature $T_1=20$~К and
gas column density $N_1=2\times10^{22}$~cm$^{-2}$,
surrounded by a warm envelope with
$T_2=400$~К and $N_2=2\times10^{14}$~cm$^{-2}$.
However, a comparably good agreement
with the observed SED is given by a model with the
opposite order of the components, i.e., with a hot
core and cold envelope, and it is impossible to choose
between these two models using this approach.

Another shortcoming of the two-component
model is its phenomenological character: we specify
layer temperatures and column densities without
considering how they could be realized physically, if
they can be realized at all. Therefore, the results of
such modeling should be interpreted with caution.
For instance, the following questions arise in connection
with the two-component model. First, if a cold
core is surrounded by a warm envelope, how was the
envelope warmed? Suppose that the cloud is warmed
by some external radiation field. If so, is it possible
to choose parameters of this field that can reproduce
the observed spectrum of the cloud? Second, if we
suppose in a model with a hot core and cold envelope
that the inner parts of the cloud are heated by some
source, e.g., a protostar, is it possible to find protostar
parameters that yield the necessary thermal structure
of the cloud and reproduce the observed SED?

We must bear in mind that this modeling took
into account only the spectrum toward the center.
However, we know both the SED toward the cloud
center (or the integrated spectra) and the distribution
of the radiation intensity, i.e., maps of the object
at the different wavelengths, from the observations.
In particular, the NIR intensity distributions for the
IRDC cores show that the intensities decrease toward
the cloud centers; i.e., these objects are seen in
absorption, in contradiction with the initial assumptions
of our two-component model. Thus, although
such simplified models are formally able to describe
the observed spectrum, they may not help us make
progress in reconstructing the physical structures of
protostellar clouds.

Therefore, it seems logical to develop more physically
feasible models of protostellar clouds and compare
them with observations in more detail. A model
of this type is described in the next subsection.

\subsection{Spatial Model of a Protostellar Cloud}

We suppose that the cloud is spherically symmetric,
and its density distribution can be described, like
those in low-mass prestellar cores~\cite{Tafalla2002},
using the expression
\begin{equation}
n(H_2) = \dfrac{n_0}{1+\left(\dfrac{r}{r_0}\right)^\beta},
\end{equation}
where $n_0$ is the central number density of hydrogen
nuclei, $r_0$ is the radius of an inner region with approximately
constant density $n$, and $\beta$ is the index for the
density decrease in the envelope. The outer radius of
the cloud is taken to be 1 pc. In the radiative transfer
modeling, we must also specify the inner radius of the
cloud, which is always taken to be 50~AU.

To take into account possible differences in the
evolutionary status, e.g., the existence of a protostar,
we assume that there is a source with radius $R_\ast$
at the center of the cloud, which emits as an ideal black body
with temperature $T_\ast$. The exact nature of this source
is unimportant: this may be the emission of either
the protostar or accreting matter. The central source
determines the temperature of the inner parts of the
cloud. The cloud is irradiated from the outside by
isotropic interstellar background emission with color
temperature $T_{\rm bg}$ and dilution $D_{\rm bg}$,
which determines the temperature in the outer parts of the cloud.

In addition, we specify the isotropic IR background
at the wavelengths used for comparison with
the observations. The intensity of this background
is equal to the observed intensity at the edge of
the cloud. At wavelengths shorter than 10 $\mu$m, the
IR background is probably due to the emission of
small interstellar dust grains, polycyclic aromatic
hydrocarbons (PAHs) \cite{draineli}. Note that the intensity
of the NIR background is higher than the intensity of
blackbody emission with temperature $T_{\rm bg}$  and dilution
$D_{\rm bg}$, and significantly affects the cloud’s appearance
in the observations.

The main agent determining the thermal structure
of the cloud is dust, which absorbs, scatters, and reemits
the continuum radiation. Therefore, we must
solve for the continuum radiative transfer in order
to find the dust temperature distribution inside the
cloud and compute the distribution of the radiation
intensity.

We used the Accelerated Lambda Iteration (ALI)
method for the radiative transfer modeling, in which
the mean radiation intensity is determined by integrating
the radiative transfer equation along randomly
chosen directions. The algorithm used is similar
to that described in~\cite{Pavlyuchenkov2004},
but applies modifications necessary for thermal radiation.
The temperature of the medium $T$ is found from the equation
of radiative equilibrium:
\begin{equation}
\int\limits_{0}^{\infty}\alpha_{\nu}J_{\nu}d\nu =
\int\limits_{0}^{\infty}\alpha_{\nu}B_{\nu}(T)d\nu,
\end{equation}
where $\alpha_{\nu}$ is the absorption coefficient,
$J_{\nu} = (4\pi)^{-1}\int\limits_{4\pi}I_{\nu}(\vec{n})d\Omega$
is the mean radiation intensity, $I_{\nu}$ is the spectral
intensity of the radiation, and $B_{\nu}$ is the Planck function.
In integrating the radiative transfer equation,
\begin{equation}
\left(\vec{n}\nabla\right)I_{\nu} = \kappa_{\nu}\left(S_{\nu} -I_{\nu}\right),
\end{equation}
scattering is taken into account in the approximation
of isotropic coherent scattering, when the source
function $S_{\nu}$ takes the form
\begin{equation}
S_{\nu}=\dfrac{\alpha_{\nu}B_{\nu}+\sigma_{\nu}J_{\nu}}{\kappa_{\nu}}.
\end{equation}
In these equations, $\kappa_{\nu}=\alpha_{\nu}+\sigma_{\nu}$
is the extinction coefficient and $\sigma_{\nu}$ the scattering
coefficient. The absorption and scattering coefficients as
functions of frequency were computed according to the Mie theory
for amorphous silicate grains using a code presented
by D. Semenov (Max-Planck Institute for Astronomy,
Heidelberg, Germany). We assumed that the
size distribution of the dust grains is described by
a power law, $f(a)\propto a^{-3.5}$ \cite{mrn}, with minimum and
maximum grain radii of 0.001 and 10 $\mu$m.

When the temperature distribution is found, we
can compute the theoretical distribution of the radiation
intensity. For this, we used the temperature of
the medium and the mean radiation intensity, which
were determined when modeling the thermal structure
of the cloud. To compare the theoretical and
observational distributions of the radiation intensity,
we convolved the theoretical distributions with the
relevant telescope beams for each wavelength.

\begin{figure}[ht]
\setcaptionmargin{5mm}
\onelinecaptionsfalse
\captionstyle{normal}
\includegraphics[scale=0.6]{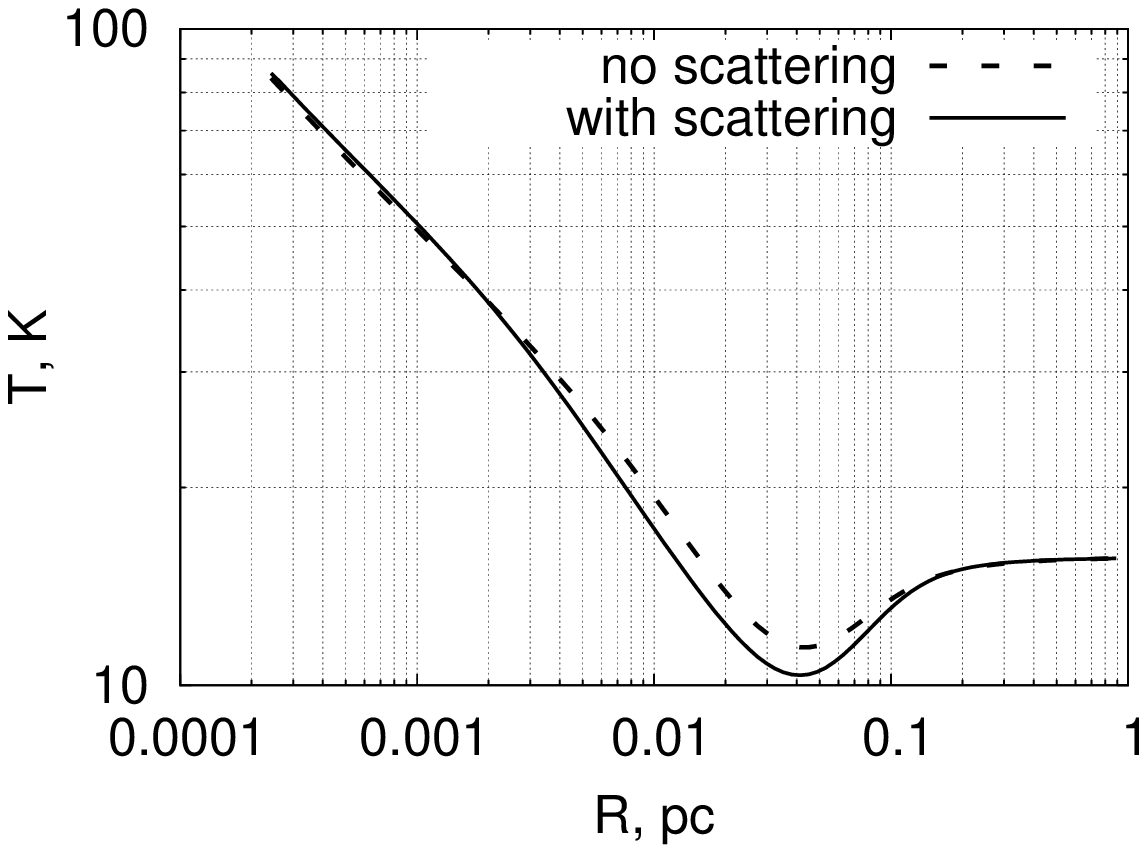}
\includegraphics[scale=0.6]{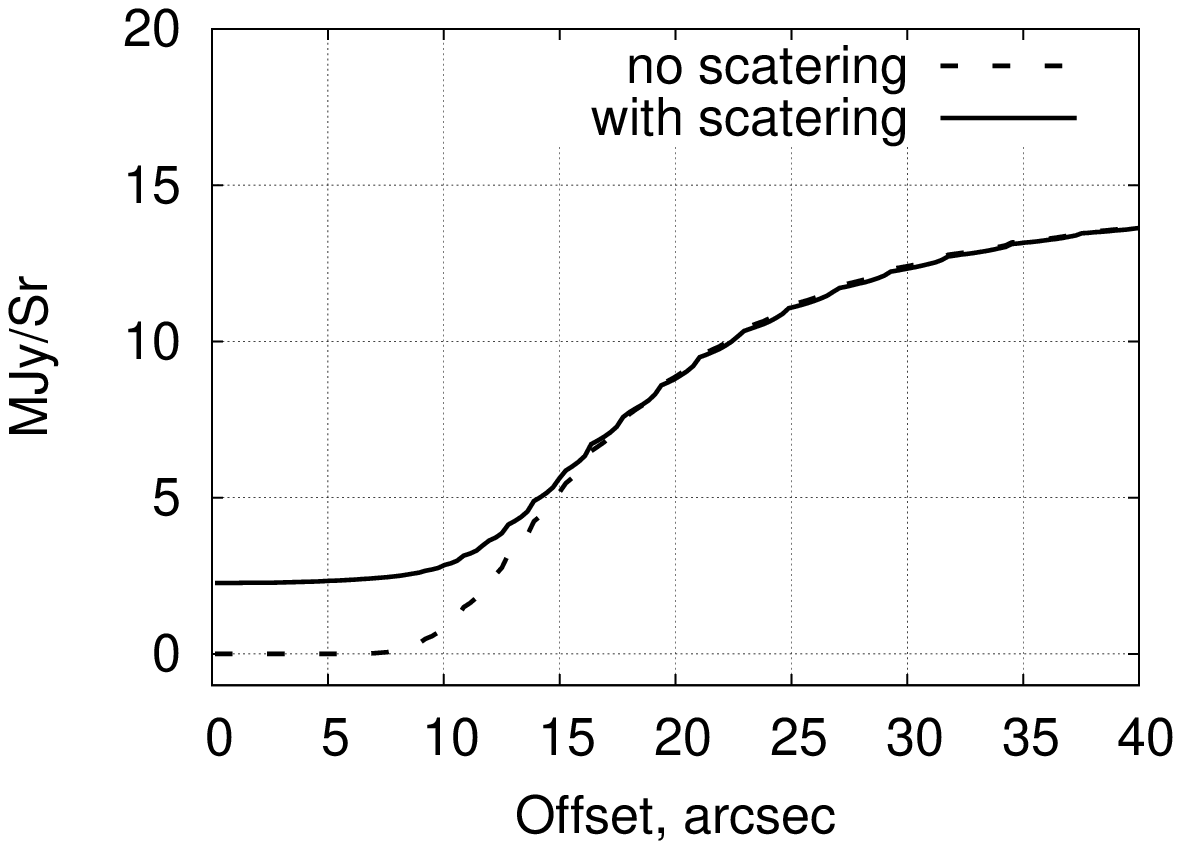}
\caption{Results of radiative transfer modeling both with and
without scattering for a representative protostellar cloud model.
The left plot shows the temperature distribution over the cloud,
and the right plot shows the intensity distribution at 24 $\mu$m.}
\label{pic1}
\end{figure}

We will now demonstrate that we must include
scattering in the radiative-transfer modeling. Figure~\ref{pic1}
shows the model temperatures and intensity distributions
at 24 $\mu$m obtained for a representative cloud
with a central density $n_0=10^7$~cm$^{-3}$ and a central
source temperature $T_{\ast}=5000$~К, both with and without
scattering. The temperature distribution in the
cloud essentially does not depend on the scattering,
but scattering significantly affects the distribution of
the emergent radiation. Without scattering, the intensity
distribution has deep minima, with the minimum
depth determined by the column density toward
the cloud center. After taking scattering into account,
the decrease in the intensity toward the cloud center
grows weaker, with the minimum intensity being determined
by the ratio of the absorption coefficient to
the scattering coefficient.

The method used to compute the temperature and
intensity distributions was tested using a number
of model problems. In particular, our computational
results on the dust temperature agree with the solution
obtained with the TRANSPHERE-1D numerical
code, which was developed by C. P. Dullemond for
modeling radiative transfer in dust envelopes taking
into account the thermal emission of the 
dust\footnote{http://www.mpia.de/homes/dullemon/radtrans}.

\section{MODELING RESULTS FOR IRDC 320 AND IRDC 321}

The method used to model the structures of protostellar
clouds and the intensity distributions was
applied to IRDC 320 and IRDC 321. We selected five
variable model parameters (Table~\ref{param}); three of these
($n_0$, $r_0$ and $\beta$) characterize the density distribution,
and the other two ($T_\ast$ and $D_{\rm bg}$) the intrinsic and external
radiation fields. The other parameters are fixed,
and are also presented in Table~\ref{param}. The inner radius
of the cloud was specified in order to approximately
take into account the absence of dust near the internal
source (due to sublimation). In principle, the dust sublimation
radius depends on the source parameters,
but we chose a fixed radius of 50~AU, which in
our calculations corresponds to the upper limit for
the size of the sublimation zones around hot stars.
If there is no star, than the inner radius of the cloud
essentially does not affect the distribution of emitted
radiation. The outer radius of the cloud, 1 pc, is equal
to the observed radii of the studied cores. The radius
of the central source (star), 5~$R_\odot$ \cite{whitney},
is chosen fairly arbitrarily, assuming that, within the explored range
of stellar parameters, the heating will depend on the
star’s luminosity, and an incorrect choice of the star’s
radius can be compensated for through a suitable
correction of its temperature. The temperature of the
interstellar radiation field was taken to be $10^4$~K.

For each parameter combination, we modeled the
radiative transfer, computed the intensity distributions,
and quantitatively compared the computed and
observed distributions. The search for the best-fit
values of the variable parameters was done with the
PIKAIA genetic algorithm. The models were compared
with observations at 1.2 mm and 70, 24 and
8~$\mu$m. As the clouds were assumed to be spherically
symmetric, the radiation intensity depends only on
the angular distance from the cloud center. When
constructing the one-dimensional observed intensity
distribution, we took the center to coincide with the
peak of the 1.2 mm emission.

\begin{figure}[!t]
\setcaptionmargin{5mm}
\onelinecaptionsfalse
\captionstyle{normal}
\includegraphics[scale=0.62]{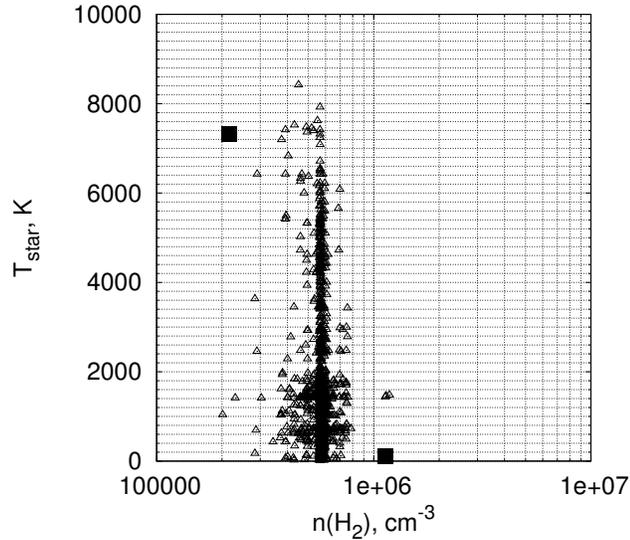}
\caption{Allowed models ($\chi^2<2.5$), obtained as a result of
fitting the theoretical intensity distribution for IRDC 320
to the observational data at 1.2 mm only.}
\label{onefit_tnh2}
\end{figure}

\begin{figure}[!t]
\setcaptionmargin{5mm}
\onelinecaptionsfalse
\captionstyle{normal}
\includegraphics[scale=0.62]{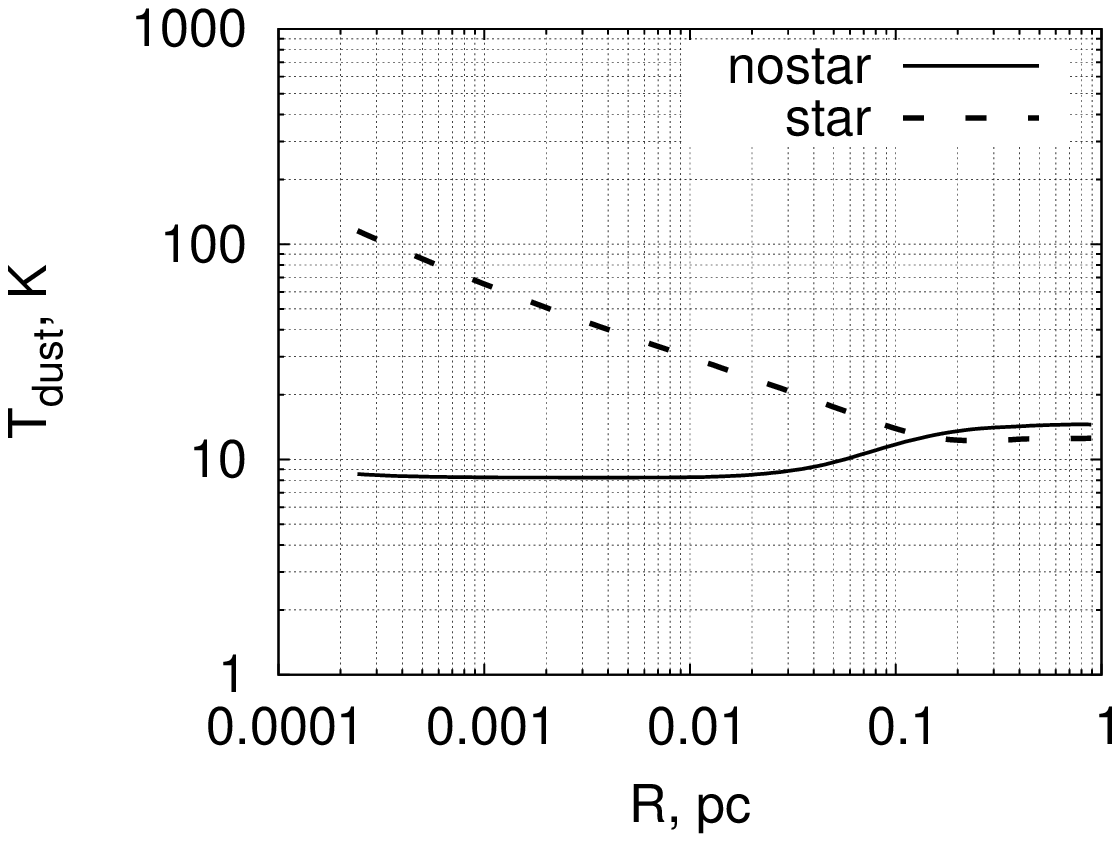}
\includegraphics[scale=0.62]{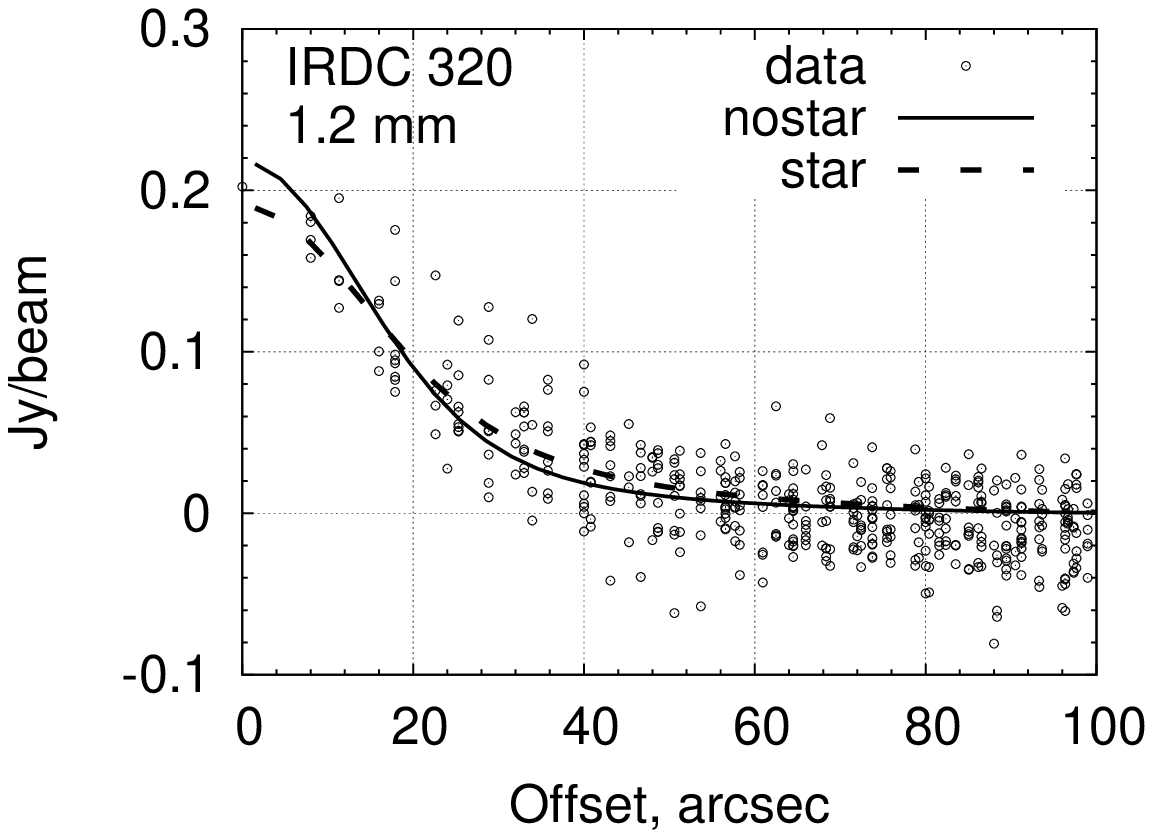}
\caption{
Comparison of the radial temperature (left) and intensity (right)
distributions for the two IRDC 320 models denoted by squares in Fig.~\ref{onefit_tnh2}. 
The dashed curve represents the model with central density $2\times10^5$~cm$^{-3}$
and central source temperature 7300 K, and the solid curve the model with central
density $1.2\times10^6$~cm$^{-3}$ and without a central source. The spectra were
fitted to the observational data at 1.2 mm only.}
\label{model2}
\end{figure}

The agreement between the observed and model
intensity distributions at each wavelength can be
characterized using the standard $\chi^2$ criterion. The
generalized agreement parameter that must be minimized
by the genetic algorithm is equal to the sum
of the $\chi^2$ values for the individual wavelengths. We
aim to determine the physical structure of the object
using observations at both IR and millimeter wavelengths.
The need to use both these wave bands is
demonstrated by the ambiguity of the results obtained
using only long-wavelength data. The modeling of
IRDC 320 using only data at 1.2 mm formally yields
the following best-fit parameters of the cloud: the
central H$_2$ density is $5.1\times10^5$~cm$^{-3}$, the plateau
radius $1.9\times10^4$ AU, the density-profile index 3.2,
and the mass 580\,$M_\odot$. However, this minimum is
very flat. Figure~\ref{onefit_tnh2} exhibits the model distribution in
the $T_\ast$--$n({\rm H}_2)$ plane. We can see that the algorithm
cannot distinguish between the models with and
without an internal source. As an illustration, we
selected two models denoted in Fig.~\ref{onefit_tnh2}
by the black squares. These are located at the edges of the range
of allowed models: a less dense core with an internal
source and a denser core without such a source.
Figure~\ref{model2} displays the distributions of the temperature
and intensity for these two models. Even with the
very different temperature distributions, the difference
between the 1.2~mm intensity profiles is negligible.
Thus, it is undesirable to use only millimeter
wavelengths when determining the temperatures and
densities of prestellar and protostellar cores.

\begin{figure}[!t]
\setcaptionmargin{5mm}
\onelinecaptionsfalse
\captionstyle{normal}
\includegraphics[scale=0.55]{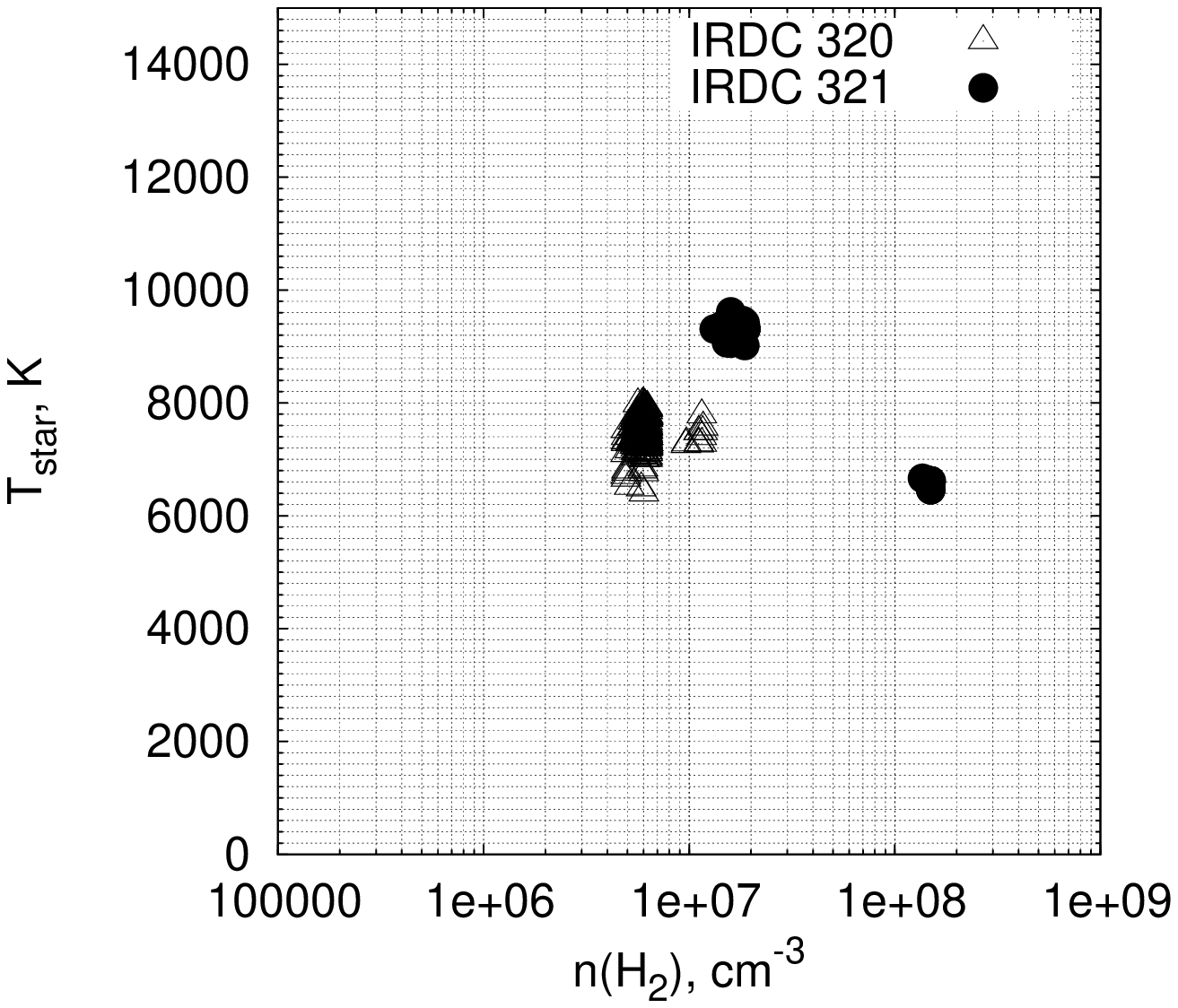}
\includegraphics[scale=0.55]{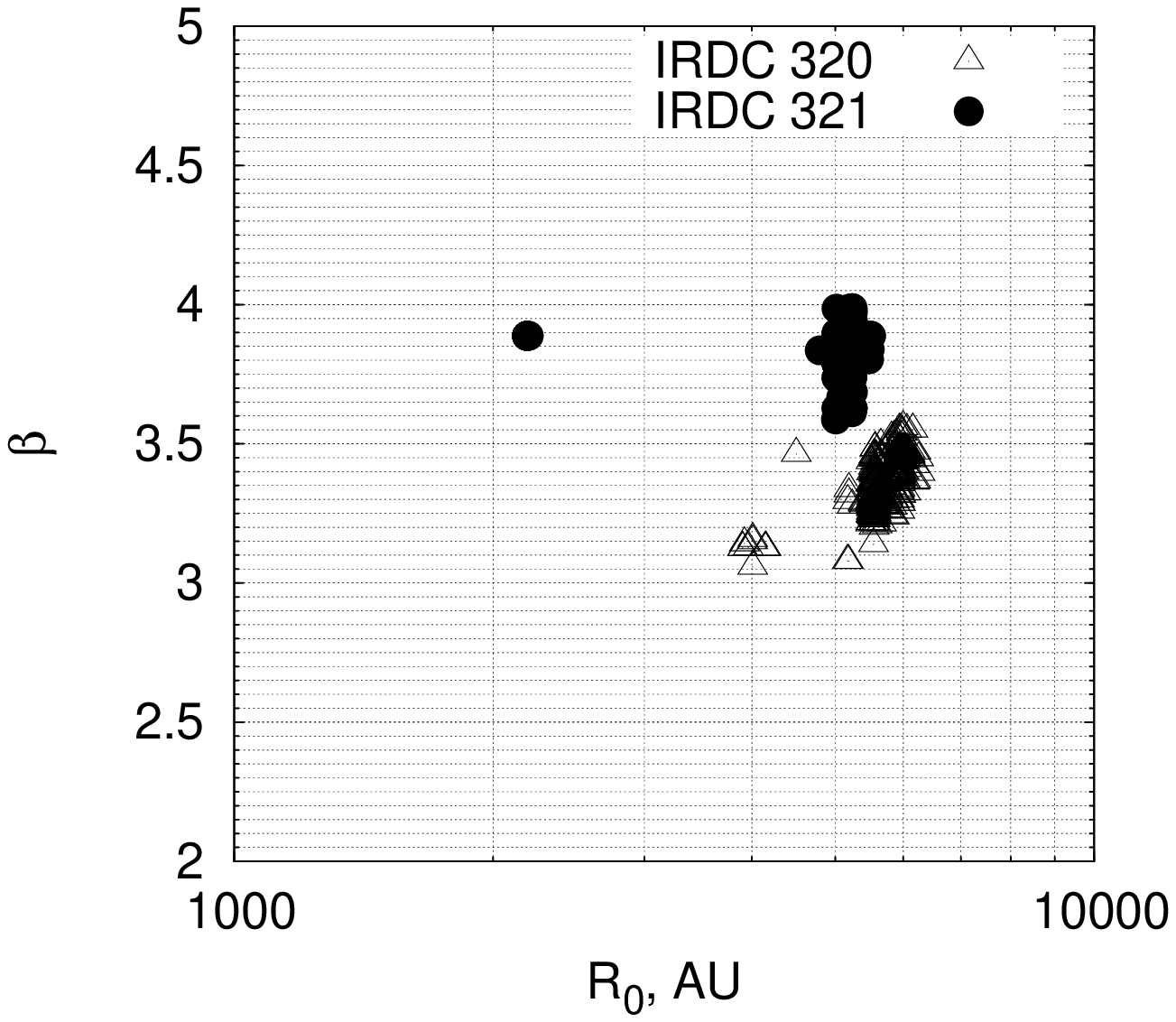}
\caption{
Localization of the allowed parameters for the IRDC 320 (circles)
and IRDC 321 (triangles) models in $n_0$ -- $T_{\ast}$ space (left)
and $R_0$ -- $\beta$ space (right)}
\label{locpar}
\end{figure}

\begin{figure}[!t]
\setcaptionmargin{5mm}
\onelinecaptionsfalse
\captionstyle{normal}
\includegraphics[scale=0.62]{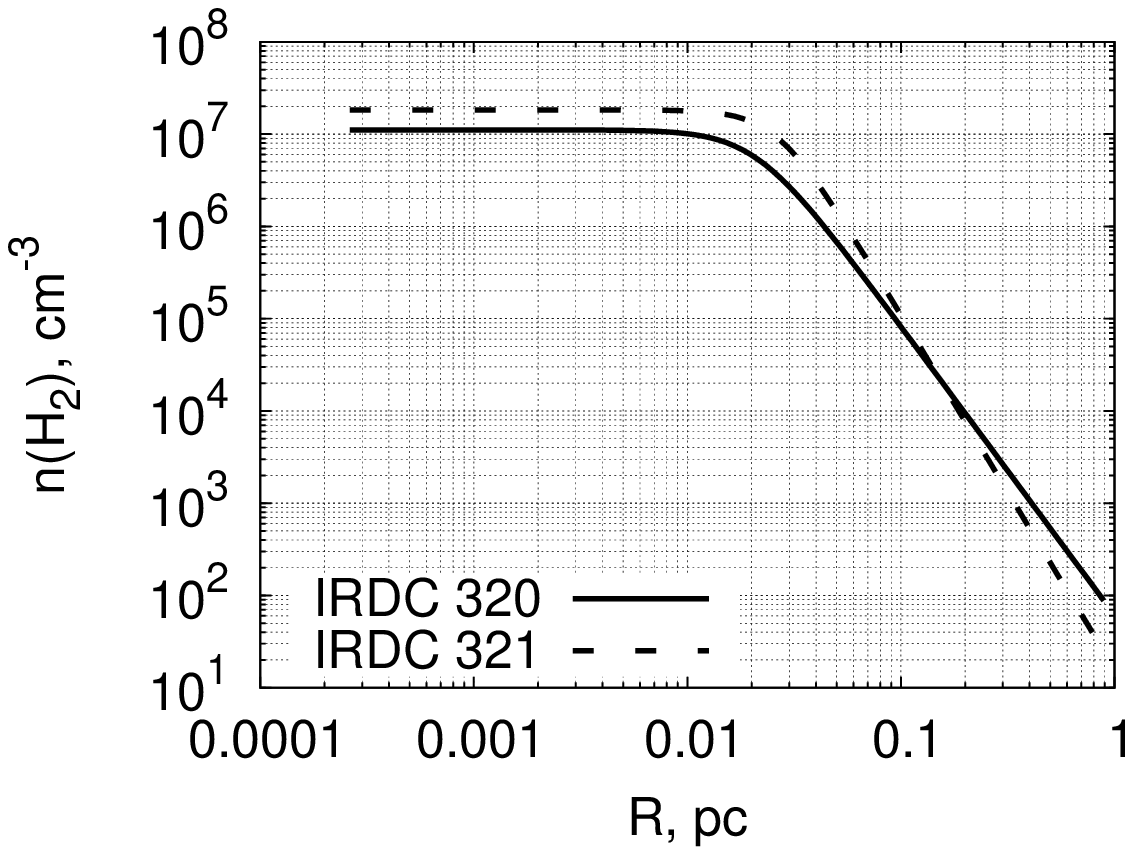}
\includegraphics[scale=0.62]{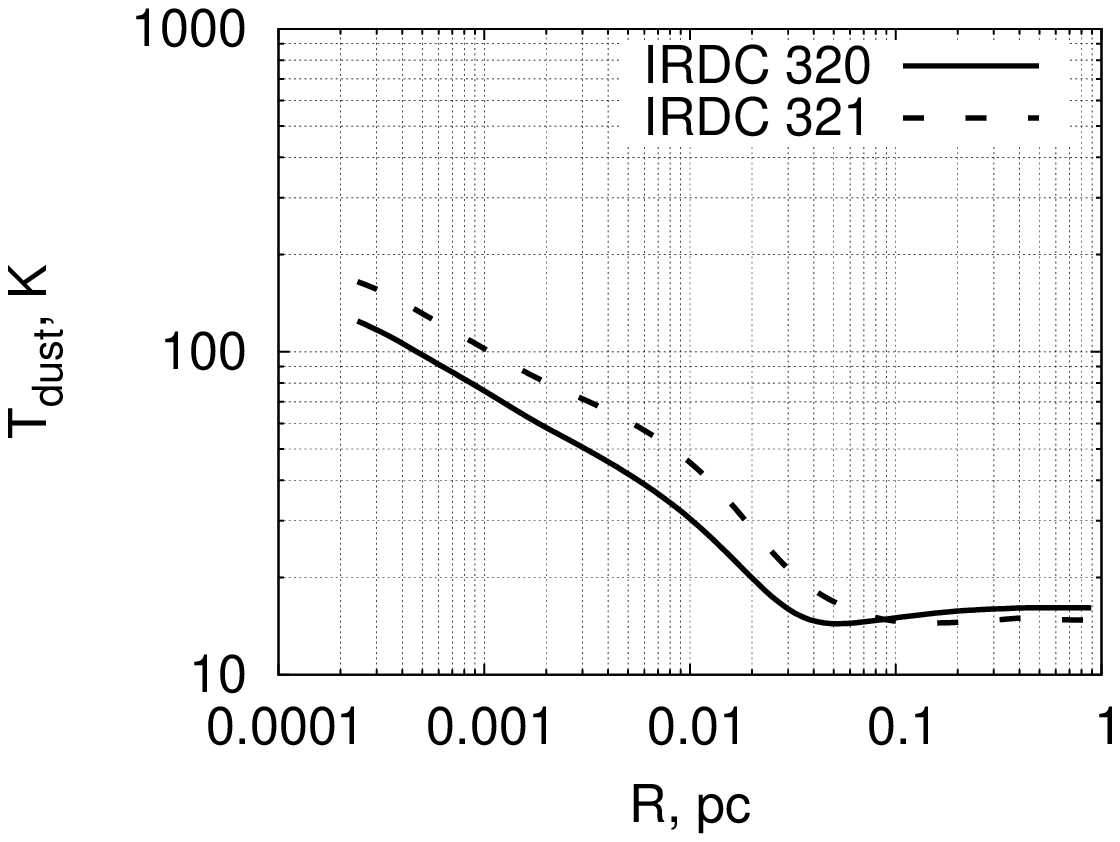}
\caption{Distributions of the density (left) and
temperature (right) in the models for IRDC 320 (solid)
and IRDC 321 (dashed).}
\label{denstemp}
\end{figure}

Further, we present our modeling results obtained
using four wavelengths. The localization of the allowed
parameters for both sources is shown in Fig.~\ref{locpar},
which displays all models with $\chi^2<12$ for IRDC 320
and $\chi^2<11$ for IRDC 321. The corresponding best-fit
parameters for IRDC 320 and IRDC 321, are
given in Table~\ref{param}, together with the masses and hydrogen
column densities toward the cloud centers. The
temperature and column-density distributions for the
best-fit models are shown in Fig.~\ref{denstemp}.

In the case of IRDC 321, the allowed models
are located in the region of higher densities, with a
concentration of the density toward the cloud center
(demonstrated by the high values of $\beta$). IRDC 320 is
characterized by flatter matter distributions and lower
central densities. However, one must recognize that
there remains some ambiguity in the solutions, even
after fitting the model SEDs at four wavelengths.

The formal best fit for IRDC 321 (when the
computation was terminated) has the central density
$1.8\times10^7$~cm$^{-3}$, central plateau radius 5000 AU, 
$\beta=3.8$, and central source temperature $T_{\ast}=9300$\,K.
However, there is another solution having nearly
the same $\chi^2$, with a higher central density
of $1.8\times10^8$~cm$^{-3}$, central plateau radius of 2200 AU, and
central source temperature of 6700 K. Thus, there remains
some ambiguity for this source, which prevents
us from distinguishing between models with denser,
compact cores and colder stars and those with less
dense, extended cores and hotter stars. Nevertheless,
neither of the best-fit models admits the absence of
a central source. In other words, the SED features
of this source, especially the existence of the 70 $\mu$m
emission, cannot be explained if it is a starless source.

The situation with IRDC 320 is slightly more definite.
At the termination of the computations, the best
agreement with the observations was achieved for the
model with central density $1.1\times10^7$~cm$^{-3}$, 
central plateau radius 4000 AU, $\beta=3.1$, and central
source temperature $T_{\ast}=7300$\,K. Other models with similar
$\chi^2$ values have approximately the same parameters;
in contrast to the situation for IRDC 321, there is
no alternative group of allowed models. Note that our
fitting of the SED of this source over four wavelengths
also indicates the existence of a central source.

To test this conclusion, we determined the best-fit
parameters for IRDC 320 excluding the heating
by the central source, and obtained a worse agreement
with the observed SED. In this case, we found
that 70 $\mu$m is the crucial wavelength for determining
whether the central source is present or not. The profiles
of the source integrated intensity at other wavelengths
can be reproduced equally well (or poorly) as
in the model with the central source.

\begin{table}
\caption{}
{\bf Parameters of the IRDC 320 and IRDC 321 models}
\label{param}
\begin{tabular}{llll}
\hline
Parameter\hspace{4cm} &Notation\hspace{1cm}&{IRDC~320}&{IRDC~321} \\
\hline
\multicolumn{4}{c}{Adjustable parameters} \\

Central density of H$_2$, cm$^{-3}$ & $n_0$        &$1.1\times 10^7$     &$1.8\times 10^7$ \\
Plateau radius, AU                  & $r_0$        &$4\times 10^3$       &$5\times 10^3$ \\
Index of density profile            & $\beta$      &3.1                 &3.8 \\
Star temperature, K                 & $T_{\ast}$   &7300                &9300 \\
Background dilution                 & $D_{\rm bg}$ &$1.3\times 10^{-13}$ &$6.7\times 10^{-14}$ \\
\\
\multicolumn{4}{c}{Derived parameters}\\
Cloud mass, $M_{\odot}$             & $M$          &170                 &230 \\
Column density, cm$^{-2}$           & $N$          &$8.1\times 10^{23}$  &$1.6\times 10^{24}$ \\
Star luminosity, $L_\odot$          & $L$          &60                  &160 \\
\\
\multicolumn{4}{c}{Fixed parameters}\\
Internal cavity radius, AU          & $R_{\rm in}$        &\multicolumn{2}{c}{50} \\
Cloud radius, pc                    & $R_{\rm out}$       &\multicolumn{2}{c}{1} \\
Star radius, $R_\odot$              & $R_{\ast}$          &\multicolumn{2}{c}{5} \\
Background temperature, K           & $T_{\rm bg}$        &\multicolumn{2}{c}{$10^4$ } \\
\hline
\end{tabular}
\end{table}

\begin{figure}[!t]
\setcaptionmargin{5mm}
\onelinecaptionsfalse
\captionstyle{normal}
\includegraphics[scale=0.6]{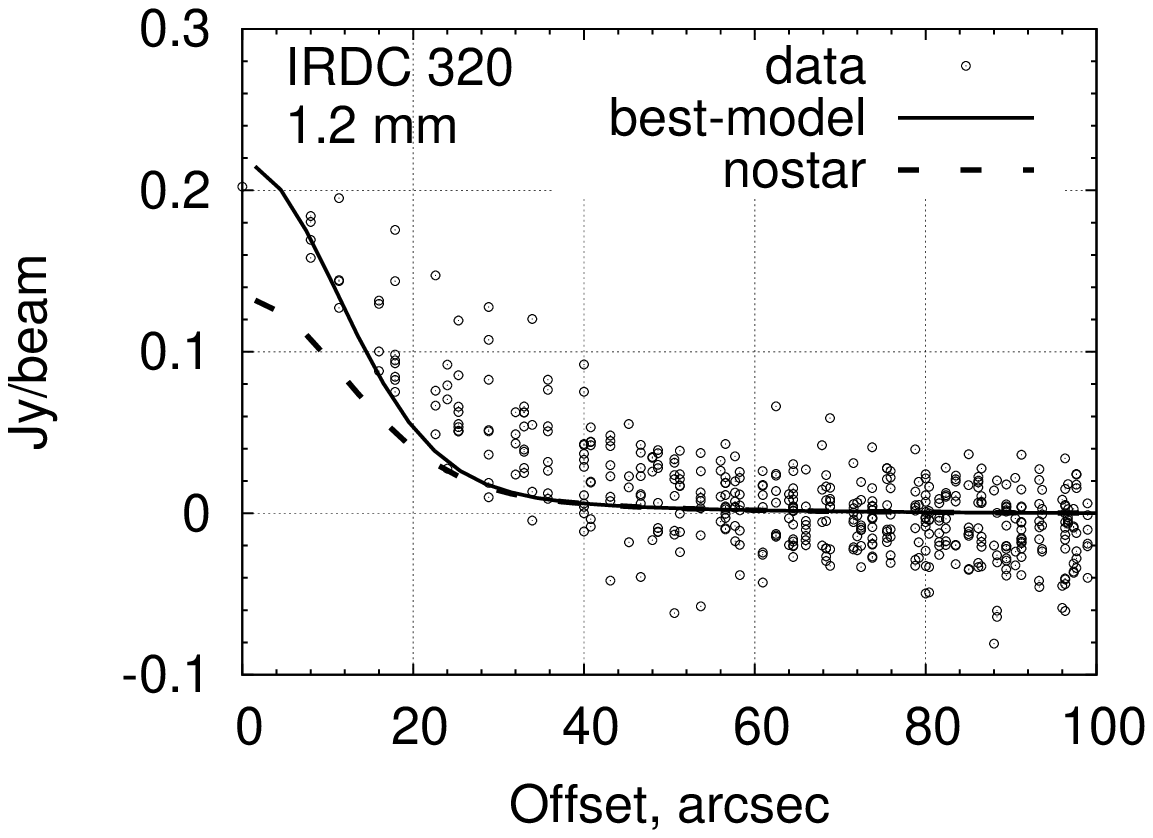}
\includegraphics[scale=0.6]{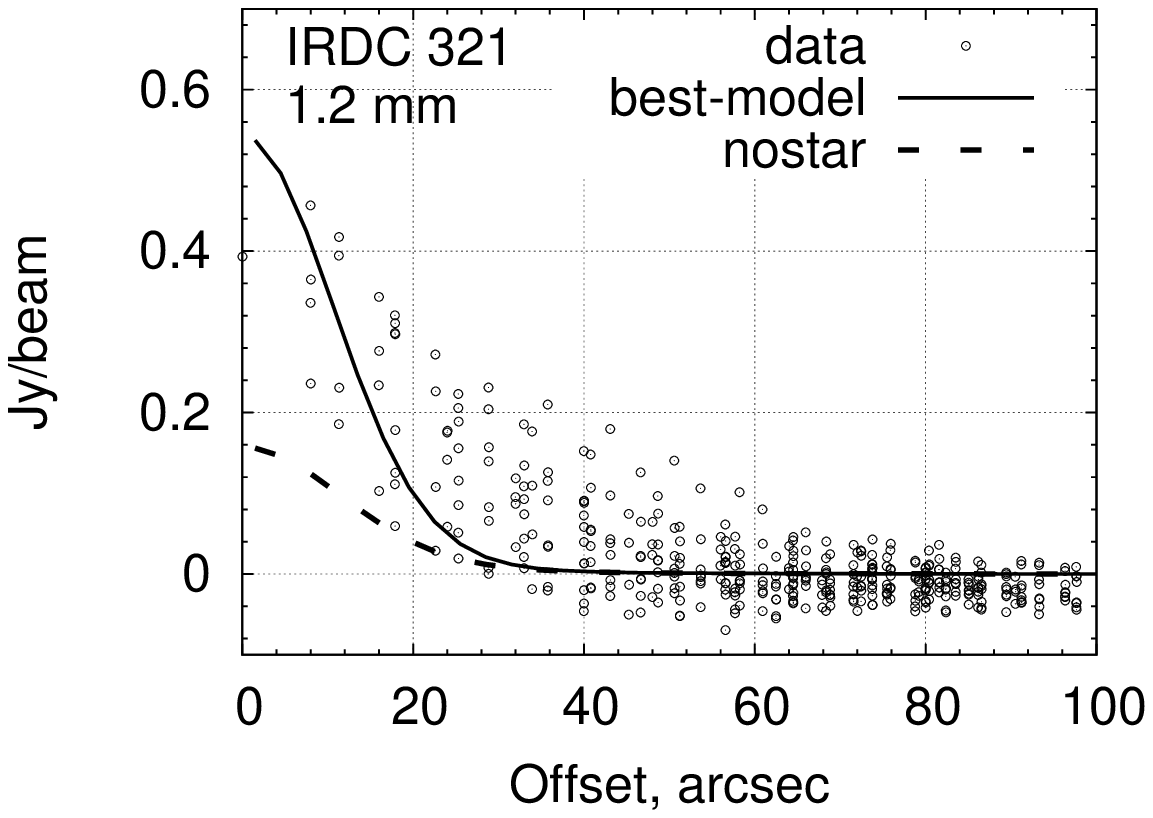}
\includegraphics[scale=0.6]{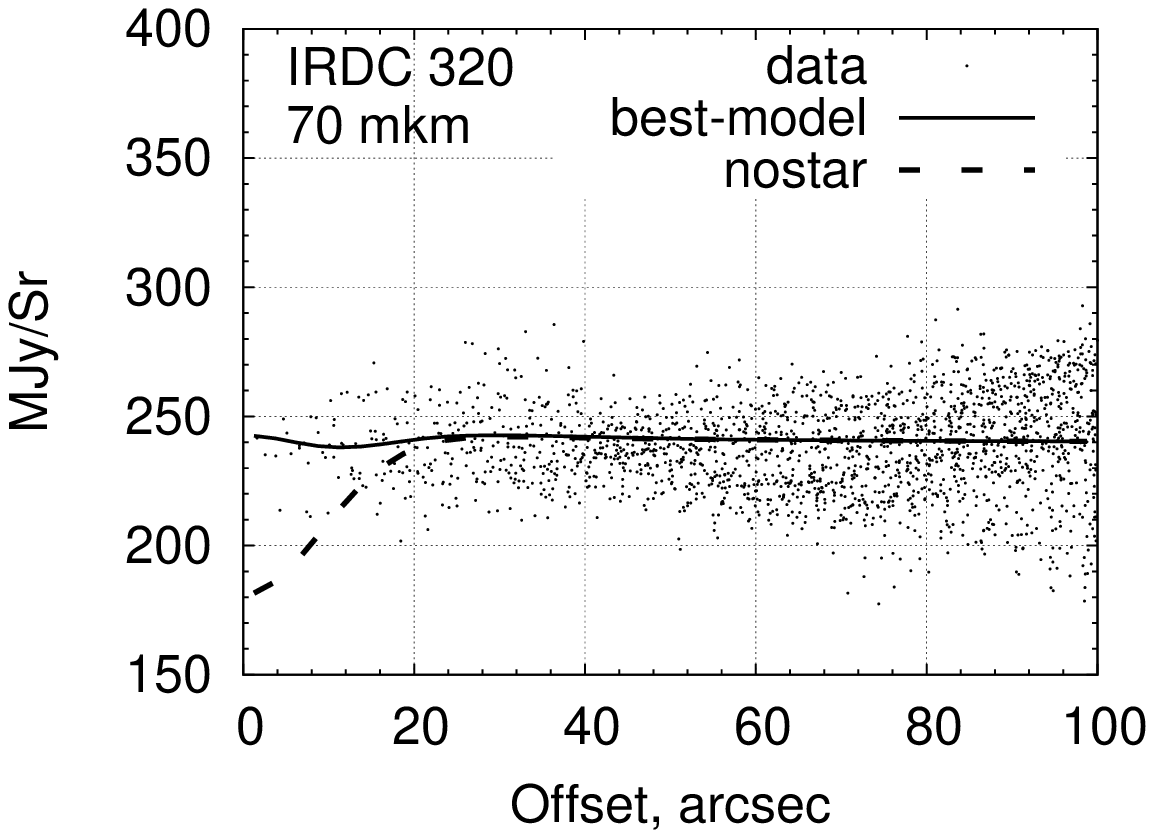}
\includegraphics[scale=0.6]{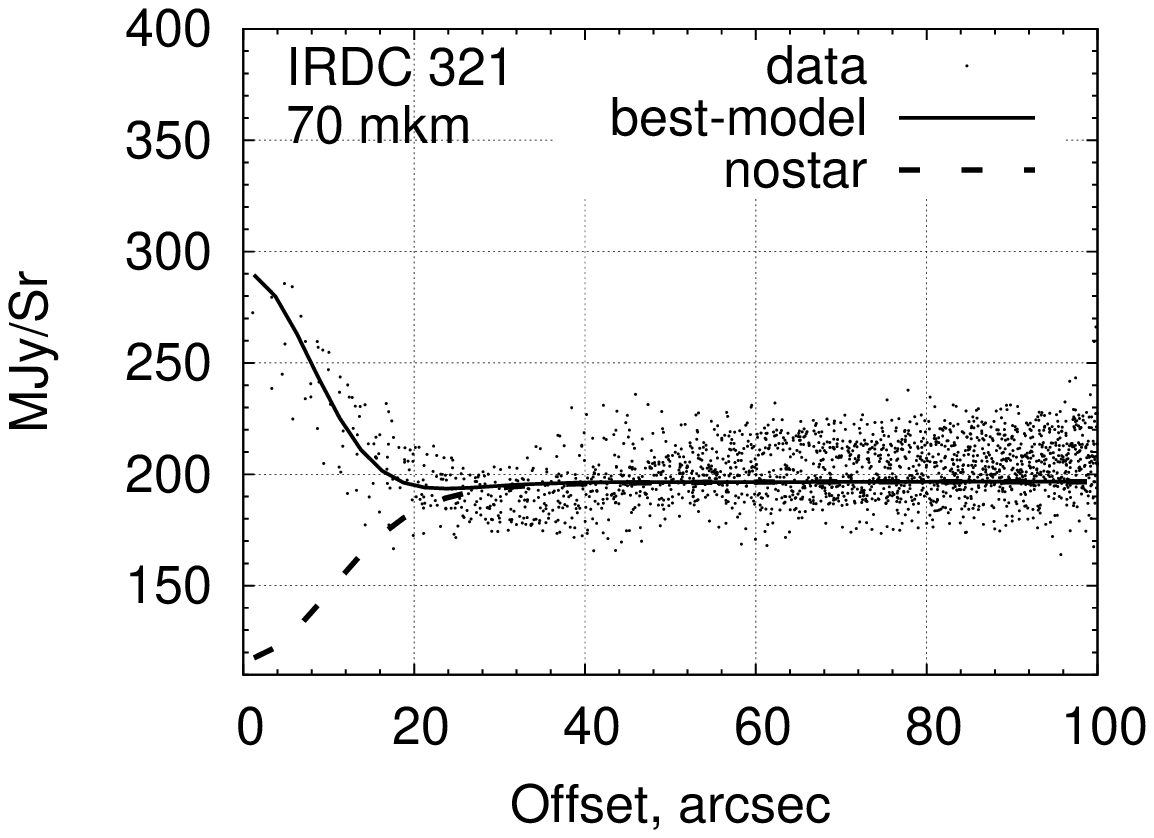}
\includegraphics[scale=0.6]{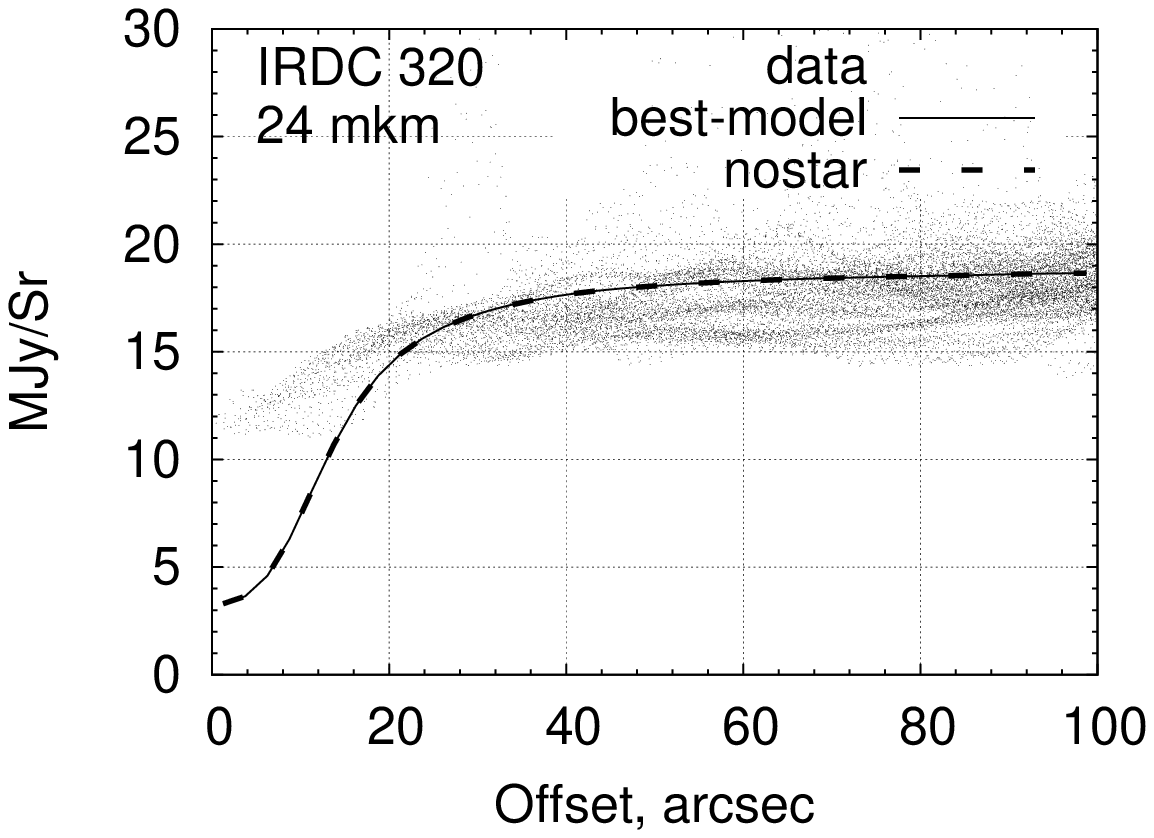}
\includegraphics[scale=0.6]{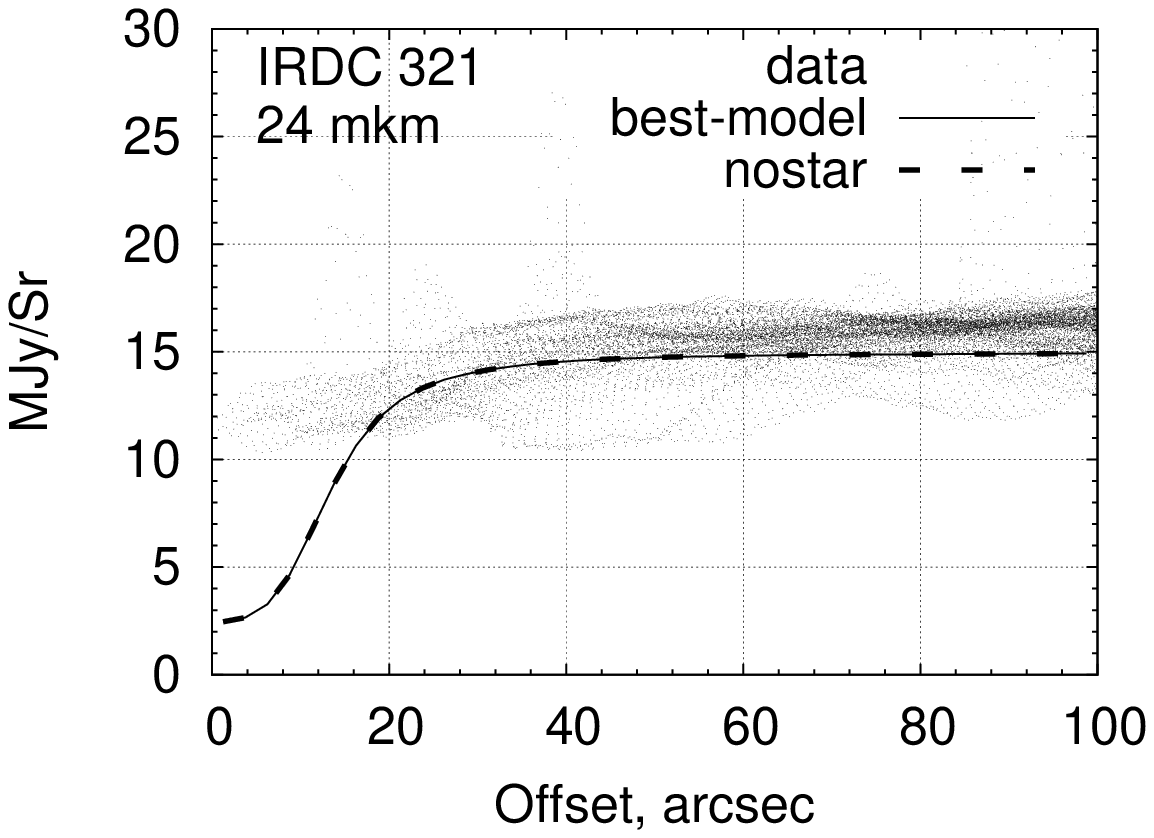}
\includegraphics[scale=0.6]{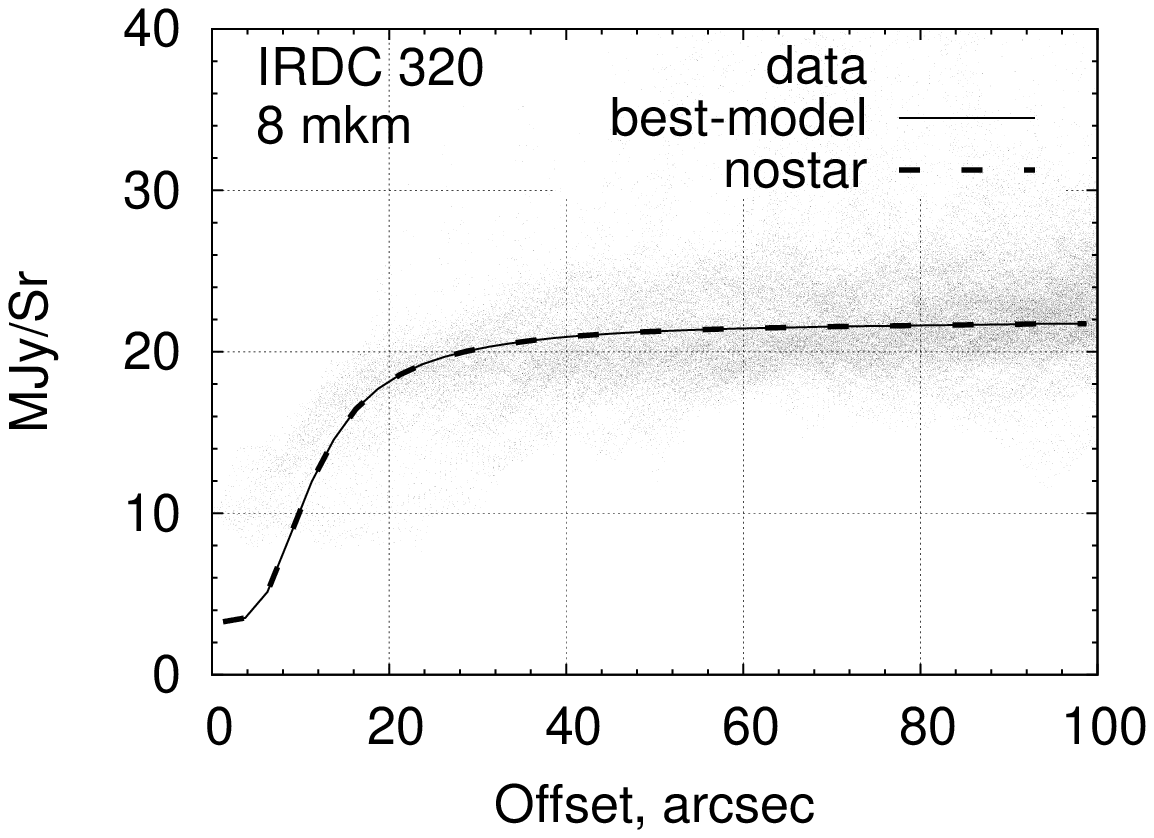}
\includegraphics[scale=0.6]{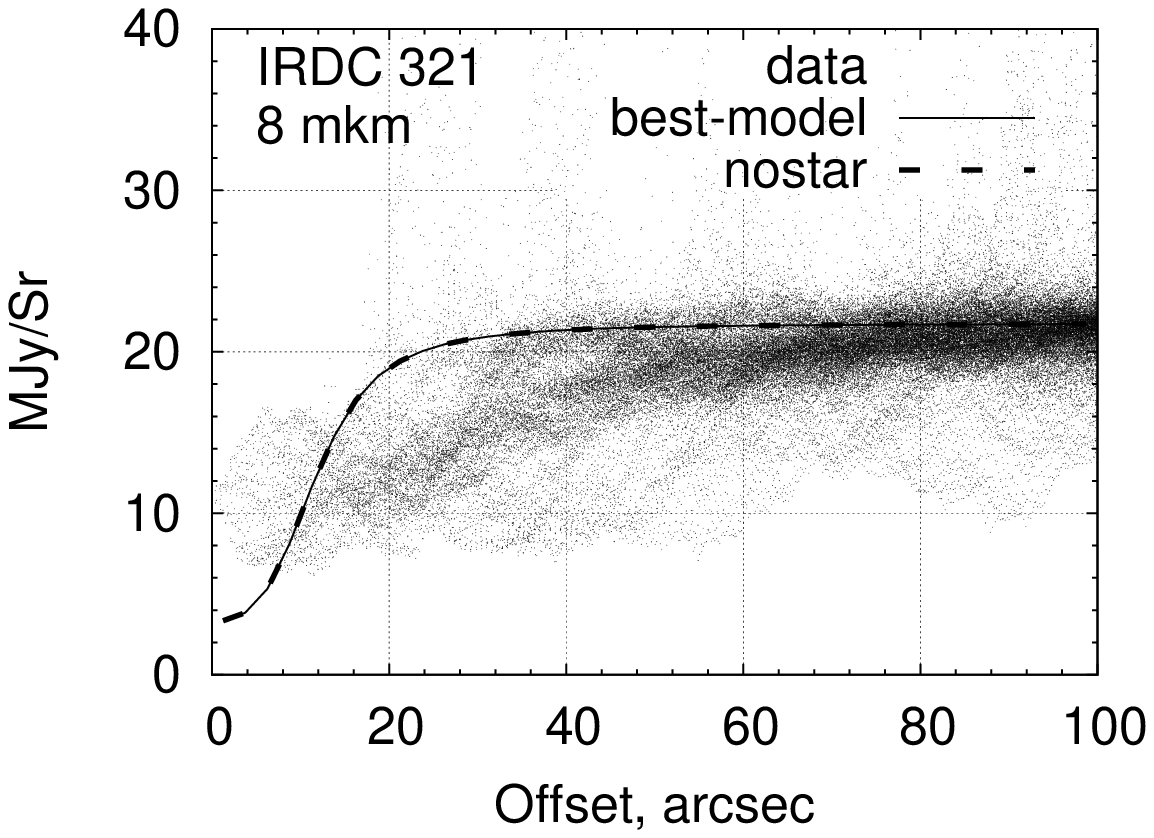}
\caption{Comparison of the model and observed intensity distributions
for IRDC 320 (left) and IRDC 321 (right). The solid curves show the
results for the best-fit models, and the dashed curves show the results
for the same models without the central sources.}
\label{pic11}
\end{figure}

Figure~\ref{pic11} compares the observed and theoretical
intensity distributions for the best-fit models. On
the whole, a good agreement was achieved for both
sources at all wavelengths. The models for IRDC 320
can explain the 1.2 mm emission, the absorption profiles
at 8 and 24 $\mu$m, and the flat intensity distribution
at 70 $\mu$m. The models for IRDC 321 yield emission
at both 1.2 mm and 70 $\mu$m, whereas absorption
profiles appear at both 8 and 24 $\mu$m. The cloud masses
are comparable, but the central density and column
density is appreciably higher in the IRDC 321 model
than the IRDC 320 model. The higher central density,
column density, and stellar temperature in the IRDC
321 model compared to those in the IRDC 320 model
lead to a higher 1.2-mm intensity and an increase in
the emission at 70 $\mu$m. The dashed curves in Fig. 9
show the intensity distributions calculated for clouds
with the same parameters but without the central
sources. The presence of a central star alters the
intensities at 1.2 mm and 70 $\mu$m, whereas the 8~$\mu$m
and 24~$\mu$m intensities are not sensitive to the presence
of this source.

\section{DISCUSSION}

We have described our method for studying the
structure of dense molecular clouds based on modeling
their SEDs at both millimeter and IR wavelengths.
A number of studies reconstructing the density and
estimating the masses of prestellar cores
have used observations of dust emission only in the
millimeter \cite{launhardt}. Since the intensity of this radiation in
the optically thin approximation is proportional to the
product of the dust column density and temperature,
the dust temperature must be known to estimate the
mass and density.

Including shorter-wavelength (IR) data makes it
possible to obtain a self-consistent reconstruction
of the density and temperature distributions. If the
studied object is seen in absorption, i.e., the contribution
of the dust thermal emission is negligible at
these wavelengths, the absorption intensity does not
depend on the dust temperature, and is determined
only by the dust surface density. On the other hand,
if the cloud temperature is high enough to generate
the intrinsic IR emission, the intensity of this radiation
places strong constraints on the dust temperature.
Supplementing these data with millimeter wavelength
data significantly narrows the range of
allowed model parameters for the cloud.

In other words, the millimeter-wavelength data
must be supplemented with shorter-wavelength
observations. However, modeling of the IR spectra
encounters difficulties. First, the shorter the wavelength,
the greater the importance of scattering. This
does not seriously affect the temperature distribution,
but significantly alters the shape of the SED and the
distribution of the integrated intensity (as was noted
above). Including the effect of scattering appreciably
increases the complexity of the models and makes
calculations more resource-intensive. Further, the
contribution of PAHs may be important at wavelengths
of several $\mu$m. To take this into account
correctly, we must make use of the parameters of
both the external UV radiation and the PAH particles,
which also increases the model complexity.

In the current study, we included scattering in
our modeling, but neglected possible PAH emission.
Even in this form, the model makes us able to successfully
reproduce a significant part of the observations
at the four wavelength ranges used. Among
shortcomings of the models, we note the poor fit
quality at 24 $\mu$m: the central intensity minima in the
models of both sources are deeper than the observed
minima. The higher observed intensity may be related
to scattering between the cloud and observer, which
is not taken into account in our model. However, efficient
scattering at 24 $\mu$m requires large dust grains,
which are unlikely to exist in the interstellar medium.
Another possible explanation is PAH emission in the
cloud envelope (due to the influence of interstellar UV
radiation). However, this explanation is also unlikely:
the strongest PAH bands are in the 7-8 $\mu$m band,
whereas our model successfully reproduces the intensity
of the 8 $\mu$m emission. An important factor
that could affect the spectrum of the outer parts of
the cloud in the NIR is the stochastic heating of
small dust grains~\cite{draineli_st}. Although including this effect
should not significantly affect the internal structure
of the cloud, it may be useful for determining the
UV radiation intensity at the cloud periphery. Taking
the foreground into account is also very important.
In particular, strictly speaking, our conclusion that
massive prestellar cores at 70 $\mu$m should be seen
in absorption is valid for the cores considered only
if all the background emission arises behind them.
This assumption may be approximately correct for the
relatively nearby clouds IRDC 320 and IRDC 321,
but the contribution of foreground emission may be
a critical component of the model in the more general
case, and should be determined together with other
parameters.

Some of the computed models successfully reproduce
the intensity profile at 24 $\mu$m, but not the
distributions at other wavelengths. These are models
with either extended, low-density cores (densities of
the order of $10^5$~cm$^{-3}$), or very compact cores with
rarefied envelopes. This may suggest a more complex
morphology of the studied clouds, in particular,
clumpiness and deviations from spherical symmetry.
Another reason for the discrepancy between
the model and the observations may be the failure of
the assumption that the central source is compact.
In reality, the cloud interiors may harbor a group
of protostars, or may undergo volume heating as a
result of gravitational collapse. We must also bear in
mind the significant uncertainty in the distances of
IRDCs, which affect the reconstruction of their geometric
parameters. The method used here can be used
to find the dust temperature distribution. However,
modeling of the gas-phase chemical processes requires
knowledge of the gas temperature rather than
the dust temperature. The gas temperature can be
directly determined from analysis of molecular line
intensities; however, in this case, we encounter an
ambiguity related to the complex chemical structures
of the studied clouds and non-LTE conditions of the
line formation. In particular, the temperatures obtained
using ammonia lines are relevant only to those
parts of the clouds where the ammonia molecules
exist. Modeling of the dust temperature allows us to
determine this quantity over the whole cloud. For the
purpose of chemical modeling, we can assume that
the gas and the dust temperatures are the same. This
assumption is valid for the dense regions of molecular
cloud cores~\cite{goldsmith2001}, but fails at their
periphery. Note, however, that the chemical composition
at the periphery is essentially determined by photo-processes,
whose rates do not depend on the temperature.

In spite of the difficulties noted above, the results
obtained demonstrate the effectiveness of our proposed
method for reconstructing the physical structures
of massive protostellar clumps. Further, we will
use this method to model the chemical structure of
these clumps and construct spectral maps, making
it possible to better establish their evolutionary status.
For this purpose, we must take into account
the decoupling of the gas and dust temperatures at
the periphery of the clumps, where gas heating due
to the photoelectric effect becomes important. The
low densities in these regions decreases the efficiency
of gas cooling due to collisions with dust grains,
thereby increasing the importance of gas cooling via
molecular-line and atomic-line radiation, when the
gas equilibrium temperature is higher than the dust
temperature.

\medskip

{\bf ACKNOWLEDGMENTS} \\
The authors are grateful to Th. Henning, H. Linz,
H. Beuther, B. M. Shustov, and M. S. Kirsanova for
useful discussions, and to the anonymous referee for
important comments. This work was supported by
the Russian Foundation for Basic Research (project
10-02-00612) and a Grant of the President of the
Russian Federation for the State Support of Young
Russian PhD Scientists (MK-4713.2009.2). This
work has made use of the NASA “Astrophysics
Data System Abstract Service”, the data obtained
in the GLIMPSE survey, a scientific program based
on the SPITZER results and financially supported
by NASA, and the NASA/IPAC Infrared Science
Archive, which is supported by the Jet Propulsion
Laboratory of the California Institute of Technology
in the framework of a NASA contract.



\begin{thebibliography}{99}

\bibitem{Zinnecker}
H. Zinnecker, H. M. Yorke, Ann. Rev. Astron. Astrophys. \textbf{45}, 481 (2007)

\bibitem{ref1}
M. Perault, A. Omont, G. Simon, {\it et al.}, Astron. Astrophys. \textbf{315}, 165 (1996)

\bibitem{ref2}
M. P. Egan, R. F. Shipman, S. D. Price, {\it et al.}, Astrophys. J. \textbf{494}, 199 (1998)


\bibitem{ref3}
J. M. Rathborne, R. Simon, J. M. Jackson, Astrophys. J. \textbf{662}, 1082 (2007)

\bibitem{launhardt}
R. Launhardt, D. Nutter, D. Ward-Thompson, {\it et al.} Astrophys. J. Suppl. \textbf{188}, 139 (2010)

\bibitem{alves}
J.F. Alves, C. J. Lada, E.A. Lada, Nature \textbf{409}, 159 (2001)

\bibitem{vasyuninaetal2009}
T. Vasyunina, H. Linz, Th. Henning, {\it et al.}, Astron. Astrophys. \textbf{499}, 149 (2009)

\bibitem{glimpse}
R. A. Benjamin, E. Churchwell, B. L. Babler, {\it et al.}, Publ. Astron. Soc. Pacific \textbf{115}, 953 (2003)

\bibitem{mipsgal}
S. J. Carey, A. Noriega-Crespo, D. R. Mizuno, {\it et al.}, Publ. Astron. Soc. Pacific \textbf{121}, 76 (2009)

\bibitem{ormel}
C. W. Ormel, R. F. Shipman, V. Ossenkopf, F. P. Helmich, Astron. Astrophys. \textbf{439}, 613 (2005)

\bibitem{pikaia}
P. Charbonneau, Astroph. J. Suppl. \textbf{101}, 309 (1995)

\bibitem{pikaia_agb}
A. Baier, F. Kerschbaum, T. Lebzelter, Astron. Astrophys. \textbf{516}, 45 (2010)

\bibitem{pikaia_proplyd}
C. Brinch, M. R. Hogerheijde, S. Richling, Astron. Astrophys. \textbf{489}, 607 (2008)

\bibitem{2layer}
H. Beuther, J. Steinacker, Astrophys. J. \textbf{656}, 85 (2007)

\bibitem{robitaille}
Th. Robitaille, B.A. Whitney, R. Indebetouw, K. Wood, P. Denzmore, Astrophys. J. Suppl. \textbf{167}, 256 (2006)

\bibitem{bochkarev} N. G. Bochkarev, Fundamentals of Physics
of Interstellar Media (URSS, Moscow, 2009) [in Russian].

\bibitem{Tafalla2002}
M. Tafalla, P. C. Myers, P.Caselli, {\it et al.}, Astrophys. J. \textbf{569}, 815 (2002)

\bibitem{draineli}
B. T. Draine, A. Li, Astrophys. J. \textbf{657}, 810 (2007)

\bibitem{Pavlyuchenkov2004}
Ya. N. Pavlyuchenkov, B. M. Shustov, Astron. Rep. \textbf{48}, 315 (2004)

\bibitem{mrn}
J. S. Mathis, W. Rumpl, K. H. Nordsieck, Astrophys. J. \textbf{217}, 425 (1977)

\bibitem{whitney}
B.A. Whitney, R. Indebetouw, J.E. Bjorkman, K.J.S. Wood, Astrophys. J. \textbf{617}, 1177 (2004)

\bibitem{draineli_st}
B. T. Draine, A. Li, Astrophys. J. \textbf{551}, 807 (2001)

\bibitem{goldsmith2001}
P. F. Goldsmith, Astrophys. J. \textbf{557}, 736 (2001)

\end{thebibliography}
\end{document}